\documentclass{aa}
\usepackage{graphicx}
\usepackage{txfonts}
\usepackage{aalongtable}
 \usepackage{newclude}
\usepackage[round]{natbib}
\bibpunct{(}{)}{;}{a}{}{,}
\graphicspath{{./images/}}

\def\kms{~\mathrm{km s^{-1}}}
\def\spose#1{\hbox to 0pt{#1\hss}}
\def\lta{\mathrel{\spose{\lower 3pt\hbox{$\mathchar"218$}}
     \raise 2.0pt\hbox{$\mathchar"13C$}}}
\def\gta{\mathrel{\spose{\lower 3pt\hbox{$\mathchar"218$}}
     \raise 2.0pt\hbox{$\mathchar"13E$}}}

\begin{document}

\title{The kinematics of the outer halo of M87\\
  as mapped by planetary nebulae\thanks{Based on observations made with the VLT at Paranal
    Observatory under programs 088.B-0288(A) and 093.B-066(A), and
    with the SUBARU Telescope under program S10A-039.}}

\author{A. Longobardi\inst{1,2} M. Arnaboldi\inst{3},
  O. Gerhard\inst{2}, C. Pulsoni\inst{2}, I. S\"oldner-Rembold\inst{2}}

\offprints{A. Longobardi}

\institute{Kavli Institute for Astronomy and Astrophysics, Peking
  University, 5 Yiheyuan Road, 100871 Beijing, China\\ e-mail:
  alongobardi@pku.edu.cn\and Max-Planck-Institut f\"ur
  Extraterrestrische Physik, Giessenbachstrasse, D-85741 Garching,
  Germany \\ e-mail: gerhard@mpe.mpg.de,
  cpulsoni@mpe.mpg.de \and European Southern Observatory,
  Karl-Schwarzschild-Strasse 2, D-85748 Garching, Germany \\ e-mail:
  marnabol@eso.org \\}

\date{Draft......, Received .......; Accepted .......}

   \authorrunning{A.Longobardi et al.}
   \titlerunning{The kinematics of the M87 outer halo}

% \abstract{}{}{}{}{} 
% 5 {} token are mandatory
 
\abstract 
% context heading ({optional) 
% {} leave it empty if necessary 
{}
% aims heading (mandatory)  
{We present a kinematic study of a sample of 298 planetary nebulas (PNs) in the outer halo of the
  central Virgo galaxy M87 (NGC~4486). The line-of-sight velocities of these PNs are used to
  identify sub-components, to measure the angular momentum content of the main M87 halo, and to
  constrain the orbital distribution of the stars at these large radii.}
 % methods heading (mandatory)
{We use Gaussian mixture modelling to statistically separate distinct velocity components and
  identify the M87 smooth halo component, its unrelaxed substructures, and the intra-cluster (IC)
  PNs.  We compute probability weighted velocity and velocity dispersion maps for the smooth halo,
  and its specific angular momentum profile ($\lambda_{\mathrm{R}}$) and velocity dispersion
  profile.}
% results heading (mandatory) 
{The classification of the PNs into smooth halo and ICPNs is supported by their different PN
  luminosity functions. Based on a K-S test, we conclude that the ICPN line-of-sight velocity
  distribution (LOSVD) is consistent with the LOSVD of the galaxies in Virgo subcluster A. The
  surface density profile of the ICPNS at $100\, \mathrm{kpc}$ radii has a shallow logarithmic slope,
  $-\alpha_{\rm ICL}\simeq -0.8$, dominating the light at the largest radii. Previous B-V colour and
  resolved star metallicity data indicate masses for the ICPN progenitor galaxies of a few
  $\times 10^8 M_\odot$.  The angular momentum-related $\lambda_{\mathrm{R}}$ profile for the smooth
  halo remains below 0.1, in the slow rotator regime, out to 135 kpc average ellipse radius (170 kpc
  major axis distance). Combining the PN velocity dispersion measurements for the M87 halo with
  literature data in the central 15 kpc, we obtain a complete velocity dispersion profile out to
  $R_{\rm avg}=135$ kpc. The $\sigma_{\rm halo}$ profile decreases from the central 400~kms$^{-1}$
  to about 270~kms$^{-1}$ at 2-10 kpc, then rises again to $\simeq 300\pm50\,\mathrm{kms}^{-1}$ at
  50-70 kpc to finally decrease sharply to $\sigma_{\mathrm{halo}} \sim 100\, \mathrm{kms}^{-1}$ at
  $R_{\rm avg}=135$ kpc. The steeply decreasing outer $\sigma_{\rm halo}$ profile and the surface
  density profile of the smooth halo can be reconciled with the circular velocity curve inferred
  from assuming hydrostatic equilibrium for the hot X-ray gas. Because this rises to
  $\mathrm{v_{c,X}\sim700}\, \mathrm{kms^{-1}}$ at 200 kpc, the orbit distribution of the smooth M87
  halo is required to change strongly from approximately isotropic within $R_{\rm avg}\sim 60$ kpc
  to very radially anisotropic at the largest distances probed.}
% conclusions heading (optional), leave it empty if necessary 
  {The extended LOSVD of the PNs in the M87 halo allows the identification of several subcomponents:
    the ICPNs, the ``crown'' accretion event, and the smooth M87 halo.  In galaxies like M87, the
    presence of these sub-components needs to be taken into account to avoid systematic biases in
    estimating the total enclosed mass.  The dynamical structure inferred from the velocity
    dispersion profile indicates that the smooth halo of M87 steepens beyond $R_{\rm avg}=60$ kpc
    and becomes strongly radially anisotropic, and that the velocity dispersion profile is
    consistent with the X-ray circular velocity curve at these radii without non-thermal pressure
    effects.  }


   \maketitle
% 
% ________________________________________________________________

\section{Introduction} 
Several studies are currently concentrating on the dramatic size
growth of passive galaxies with redshift
\citep{vDokkum2010,cimatti2012} with the goal of establishing the
structural analogs in local massive galaxies. Within the cosmological
framework, a variety of different models have been put forward to
explain the mass/size growth \citep[see,][]{Huang13}. Among those, the
two phase-formation scenario appears in best agreement with
observational constraints. In this scenario, the innermost region of
massive galaxies formed the majority of their stars at $z\le 3$ on
short time-scales \citep{thomas05}, while the stars in the outermost
regions were accreted at later epochs as a consequence of mostly dry
mergers or accretion events \citep{oser10,oser12,cook16}, Then the
outermost regions of local massive galaxies should contain the fossil
records of the accretions events in form of spatial and kinematic
substructures, because the growth is expected to occur at
comparatively low redshifts and the dynamical time-scales are long
\citep{bullock05}.

In the local universe, massive galaxies are found in the densest
regions of galaxy clusters, hence a fraction of their stars in their
extreme outer regions might in fact be part of the intra-cluster light
(ICL), i.e., a stellar component that is not gravitationally bound to
a single galaxy, but orbits in the cluster potential. A galaxy's halo
and the ICL both result from hierarchical accretion; however, they
differ in their kinematics and their different levels of dynamical
relaxation \citep{dolag10,longobardi15a,cooper14, barbosa18}. Analysis
of the radius vs. line-of-sight velocity (LOSV) projected phase-space
for several massive nearby galaxies, e.g. NGC~1399
\citep{schuberth10,mcneil10} and M87 \citep{longobardi15a}, found that
halo and ICL need to be treated separately, in order to avoid
systematic biases in the mass estimates at large radii.

In massive early-type galaxies, surface brightness profiles are a
possible avenue to disentangle multiple components, those generated by
early dissipative processes or late epoch accretions, or the
ICL. The presence of an accreted component is usually inferred from
the change of slope of the galaxy's light profile at large radii
\citep{zibetti05,gonzales07,dsouza14,spavone17}, from high Sersic
indices \citep[$n>4$,][]{kormendy09}, and/or from variations of the
ellipticity profile \citep{vandokkum11,dsouza14,mihos17}. One open
question is whether the decomposition of the surface brightness
profile in multiple components is supported independently by the
stellar kinematics
\citep{hernquist91,hoffman10,emsellem04,emsellem14}. In the
interesting case of NGC~6166 the best Sersic fit decomposition of the
surface brightness profile fails to reproduce the transition between
the low velocity dispersion of the central region and the
kinematically hotter envelope at radii larger than 10 kpc
\citep{bender14}.

Therefore a kinematic decomposition is required to unambiguously
resolve the physical components. At large radii where the galaxy
surface brightness is too low for standard absorption line
spectroscopy, this can only be done with discrete tracers, such as
globular clusters (GCs) and planetary nebulas (PNs). If we were able
to measure surface brightness profiles and kinematics at large radii,
we would 1) find the predicted accretion structures where they have
not phase mixed yet; 2) isolate the kinematics of the phase-mixed,
smooth halo, to constrain its orbit distribution and obtain unbiased
estimates of enclosed mass; and 3) understand the transition between
halo and ICL.

Recent surveys of bright, discrete probes such as
GCs and PNs have enabled the systematic
studies of the physical properties of early-type galaxy halos. GCs are
compact, bright sources easily identified on high-resolution images
\citep{cote01,schuberth10,strader11,romanowsky12,pota13}. PNs, because
of the strong [OIII]$\lambda 5007\AA$ emission line from their
envelope re-emitting up to $\sim$$15\%$
of the UV-luminosity of the central star \citep{dopita92}, have
been the targets of several surveys, in order to trace light and
motions of {\it single stars} in nearby galaxies and clusters
\citep{hui93, mendez01, peng04, coccato09, mcneil10, mcNeil12,
  cortesi13, longobardi13, longobardi15a, longobardi15b, hartke17,
  pulsoni18}, and out to distances of 50-100 Mpc
\citep{gerhard05,ventimiglia11}.

The Virgo cluster, the nearest large scale structure, and its central
galaxy M87 are prime targets to address the subject of galaxy
evolution in clusters. The Virgo cluster shows a number of spatial and
kinematic substructures, with different subgroups having different
mixtures of morphological galaxy types~\citep{binggeli87}. The
evidence that many galaxies are presently in-falling towards the
cluster core, and the presence of a complex network of extended tidal
features revealed by deep photometric surveys suggest that the Virgo
cluster core is not yet in dynamical equilibrium.

The giant elliptical galaxy M87 is close to the dynamical center of
the Virgo cluster \citep{binggeli87,nulsen95,mei07} . It is classified
as a cD-galaxy, well described by a single Sersic fit with $n\sim11$
\citep{kormendy09,janowiecki10} and an extended halo that reaches out
to $R \sim$150-200 kpc. Its total stellar mass is estimated to be
$M\sim10^{12}M_{\odot}$.  The dynamical structure of M87 is dominated by random
motions, without significant rotation
\citep{vandermarel94,sembach96,gebhardt11}. A low-amplitude
kinematically distinct core \citep{emsellem14}, a slow rotational
component \citep{murphy11,emsellem14}, and a rising stellar velocity
dispersion profile with radius \citep{murphy11} were measured in
recent studies. Several independent tracers were used to probe M87's
mass distribution: X-ray measurements \citep{nulsen95,churazov10},
integrated stellar kinematics \citep{murphy11,murphy14}, GC kinematics
\citep{cote01,strader11,romanowsky12,zhu14}, and PN kinematics
\citep{arnaboldi04,doherty09}. All of these studies showed
consistently that M87 is one of the most massive galaxies in the local
Universe, but there are considerable variations among studies using
different tracers.

Of particular interest is whether the hot (T $\sim$ 1 keV) low density
(n $<$ 0.1 cm$^{-3}$) X-ray envelope \citep{forman85} around M87 is
quiescent enough to assume hydrostatic equilibrium. In this case, one
can use the temperature and density profiles derived from the X-ray
spectra to obtain the cumulative mass profile and gravitational
potential, and then estimate the orbital anisotropy of the stars from
their dispersion profile. Non-thermal contributions to the pressure
measured from X-ray data can be studied by comparing the potential
inferred from the X-rays with mass estimates from stellar kinematics
\citet[e.g.,][]{Churazov08, churazov10}.

The goals of this paper are to identify the PNs in the smooth M87 halo,
using accurate velocities and following the approach of
\cite{longobardi15a}. We work with the sample of 253 PNs M87 halo
PNs\footnote{ Compared to \citet{longobardi15a} (254), one
    repeated object has been discarded.} and the 45 intracluster (IC)
PNs, for which line-of-sight velocities (LOSVs) are available with an
estimated median velocity accuracy of $\rm 4.2\,\mathrm{kms^{-1}}$. 
We dermine the rotation $v$ and velocity dispersion $\sigma$ for the
M87 halo in the region from $\sim$ 20 kpc to $\sim$ 200 kpc.  From
these measured profile we aim to answer the question whether the halo
stellar population, its mean square velocity, and its degree of
relaxation change smoothly with radius, such that it eventually
reaches ICL properties, or whether the halo and ICL are distinct
populations and dynamical components. Furthermore, we will investigate
whether the halo dispersion profile is consistent with the mass
profile inferred from X-rays and what this tells us about the orbital
anisotropy and its variations in the outer halo region.

The paper is structured as follows: in Sect.\ref{sec2} we revisit the
PN LOSVD and re-identify M87 halo PNs and ICPNs. In Sect.\ref{sec3} we
estimate the smoothed velocity field for the M87 halo PNs, and
determine the amplitude of rotation and the $\lambda(R)$ angular
momentum parameter as a function of radius. In Sect.~\ref{sec_sigma_R}
we determine our fiducial composite velocity dispersion profile for
M87 and derive the circular velocity curve from a simple Jeans
model. This circular velocity curve is then compared with that
measured from X-ray observations in \citet{churazov10}. Finally, we
discuss our results in Sect.~\ref{discussion} and give our conclusions
in Sect.~\ref{conclusions}. In the Appendix, we provide the detailed
PN LOSVDs for the outer halo of M87 in different radial bins and the
complete M87 PN catalogue from the Suprime-Cam@Subaru photometric and
the FLAMES@VLT spectroscopic PN surveys.

In this work, the systemic velocity of M87 is
$\rm{V_{sys}=1307.0\pm 7\, km s^{-1}}$ \citep{allison14}, and we adopt
a distance modulus of 30.8 for M87 \citep{longobardi15a}, implying a
physical scale of 73 pc arcsec$^{-1}$.

%SECTION 2

\section{The Kinematics of M87 and the Virgo intracluster stars}% and unrelaxed components} 
\label{sec2}
%\subsection{Identification of the M87 halo and ICL components in the line-of-sight velocity distribution}
%\label{sec:2.1}

We begin our investigation on the kinematics of halo and IC PNs by adopting a similar approach to
\citet{longobardi15a} who applied a robust sigma estimator \citep{mcneil10} to separate the
  asymmetric broad wings of the LOSVD - the ICL- from the nearly-symmetric main distribution of
  velocities centred at the systemic velocity of M87 - the galaxy halo.

\begin{figure*}[!ht]
\centering
\vspace{+3cm}
\hspace{-1cm}

%\includegraphics[width=8.cm, height=5cm]{Pspace_halo_ICL_runningAve_nocenter_november.ps}
%\hspace{+1cm}
%\includegraphics[width=8.cm, height=5cm]{Pspace_halo_ICL_runningAve_nocenter_zomm_november.ps}
\hspace{-0.5cm}
\includegraphics[width=6.cm]{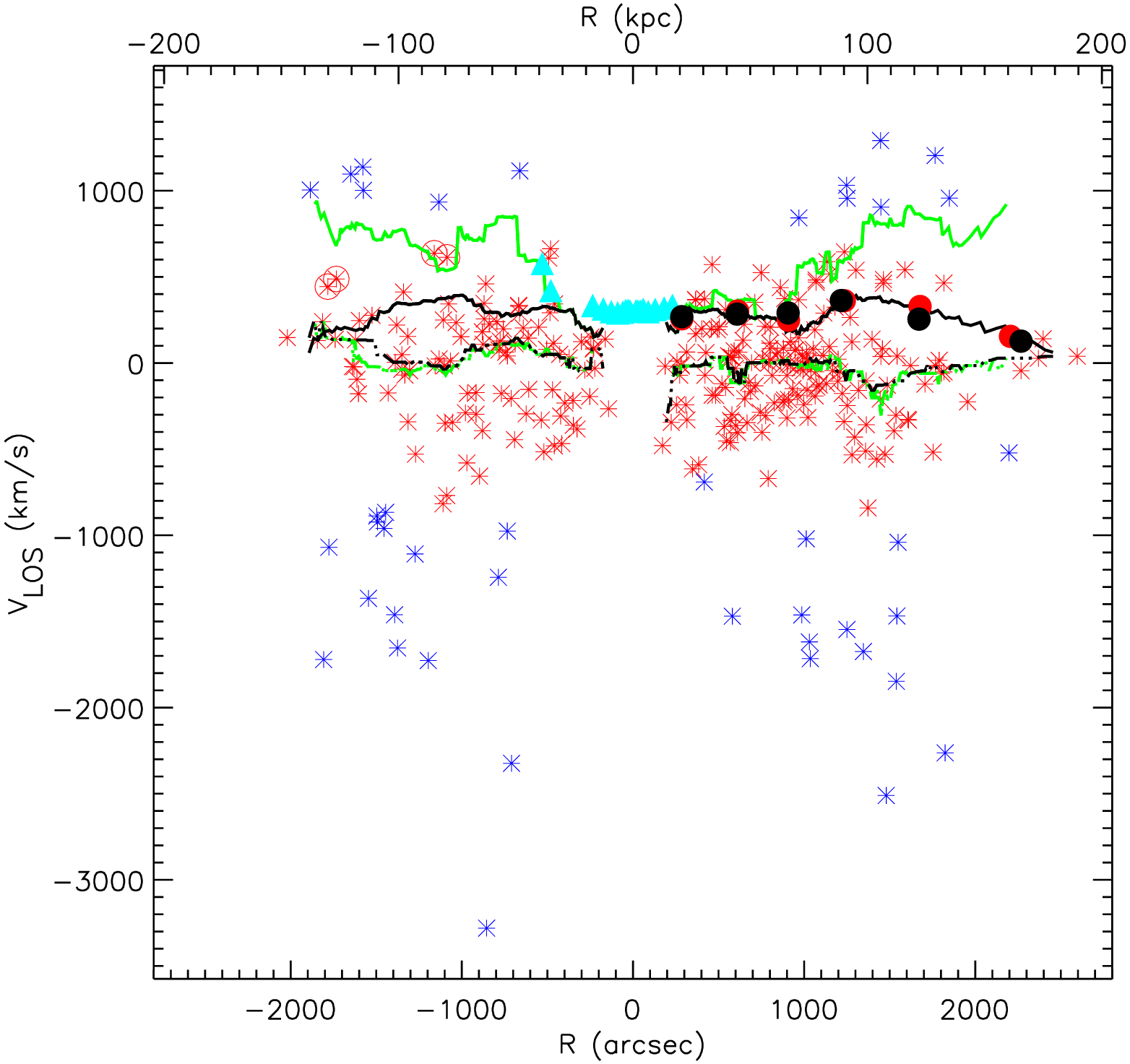}
\hspace{0.5cm}
\includegraphics[width=6.cm]{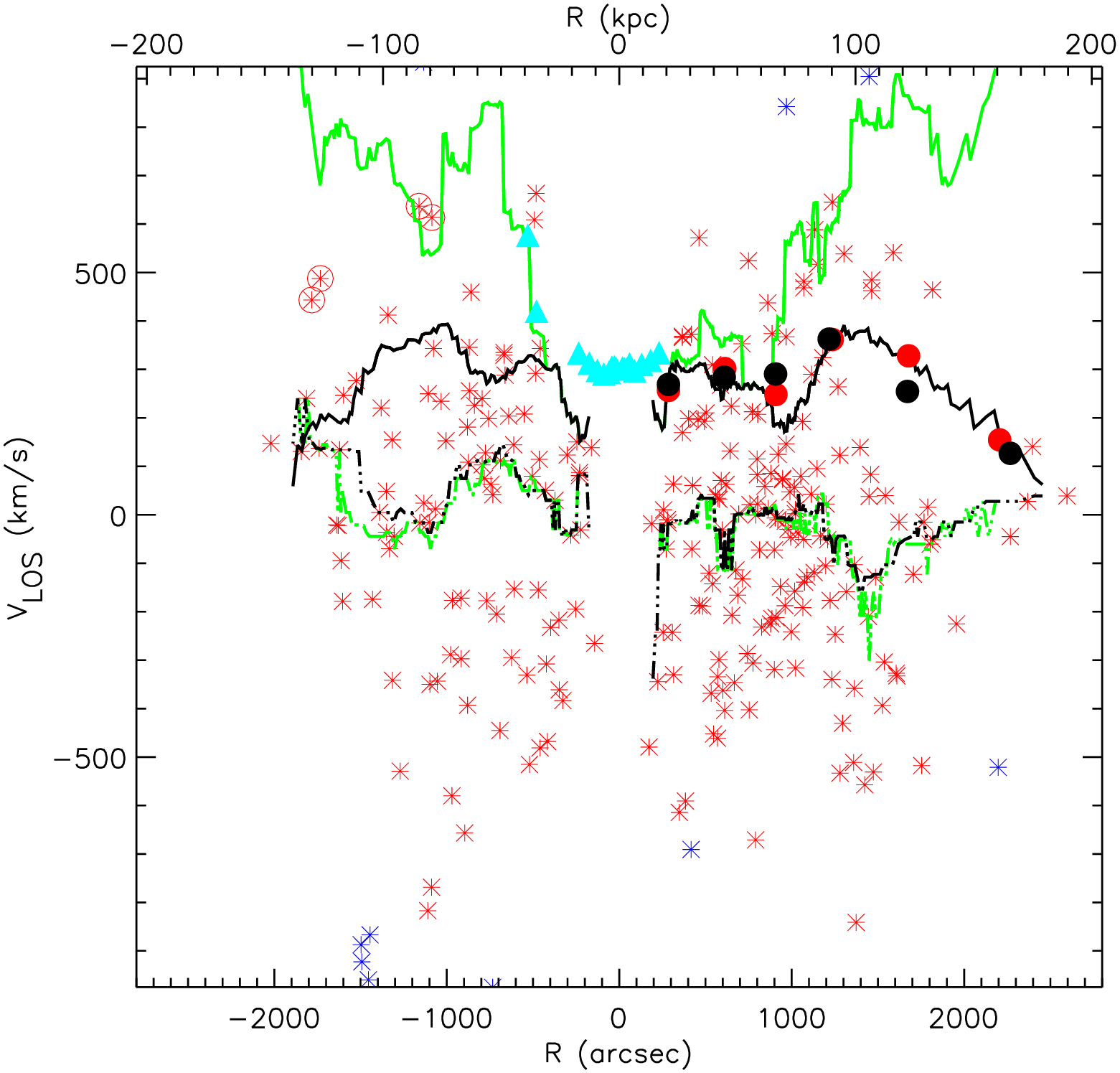}
   \caption{\small{\textbf{Left-Panel:} Projected phase-space diagram $\mathrm{V}_{\mathrm{LOS}}$
       vs.~major axis distance $R$ from the centre of M87 out to 200 kpc, for all spectroscopically
       confirmed PNs from \citet{longobardi15a}. M87 halo PNs (red asterisks) and ICPNs (blue
       asterisks) as classified by them are shown separately; $R > 0$ and $R< 0$ represent the
       northern and southern halves with respect the galaxy's center. Red open circles show four
       newly identified kinematic outliers, see Sect.\ref{outliers}. Black dashed-dot and black
       continuous lines depict the running average and running velocity dispersion independently for
       the M87 South-East and North-West, computed from the 253 halo PNs. Full red circles show the
       robust estimate of the halo velocity dispersion following \citet{longobardi15a}, while full
       black circles show the velocity dispersion values after probability-weighted removal of the
       crown substructure from \citet{longobardi15b}. Both velocity dispersion profiles (red and
       black circles) show strong radial variation, with a steep decline at major axis distances $R
       > 100 \, \rm{kpc}$.  Green dashed-dot and green continuous lines show the running average and
       velocity dispersion for the {\sl combined} sample of 253 halo PNs and 45 ICPNs; the inclusion
       of ICPNs leads to a rapidly rising velocity dispersion profile. Cyan triangles show the
       velocity dispersion measurements from the IFU VIRUS-P data of \citet{murphy11,murphy14} whose
       outer rise near $\rm{R=700\arcsec}$ ($\sim$ 40 kpc) is matched by the running joint halo and
       ICPN velocity dispersion (green line) indicative of the contribution from the
       ICL. \textbf{Right-Panel}: Zoomed-in plot in the velocity range $\pm$ 1000
       $\mathrm{kms^{-1}}$ centred on the systemic velocity of M87. }} \label{moving_sigma}
 \end{figure*}

 The halo-ICL dichotomy is illustrated in Fig.~\ref{moving_sigma} where we show the projected
 phase-space distribution for halo (red asterisks) and IC (blue asterisks) PNs,
 $\mathrm{V}_\mathrm{LOS}$ versus major axis distance, on the basis of the classification by
 \citet{longobardi15a}. Because the two components overlap in velocity, \citet{longobardi15a} argued
 that a fraction of the PNs whose LOSV values are in the range of the M87 main halo may also be IC
 PNs; this is investigated further in Section~\ref{residualICL}.

To illustrate the effect of the ICL on the velocity dispersion profile, we compute the LOSVD
\textit{running average} and \textit{running dispersion}\footnote{These quantities represent the
  mean velocity and velocity dispersion of sub-sequences of $n$ adjacent PN velocities along the
  major axis. Here $n=30$.}, for the total PNs sample (M87 halo plus ICPNs - green lines) and for
the halo PNs only (black lines), independently on both sides of the M87 major axis. While the mean
velocity curves are very similar, the running dispersion curves are widely different. The {\it
  running dispersion} of the M87 plus IC PNs (green line) quickly rises to a value which is similar
to the velocity dispersion of the Virgo sub-cluster A / Virgo core $\sim 700\, \rm{kms^{-1}}$
\citep{binggeli87,conselice01}, with no further radial variation on both sides of the M87 major
axis.  The {\it running dispersion} of only the M87 halo PNs behaves differently: it is about
constant at $ 270 - 290\, \rm{kms^{-1}}$ out to 30-40 kpc and then shows a rise at about 70 kpc,
followed by a steep decline.

We emphasise that the strong radial variations of the running dispersion for the M87 halo PNs' LOSV
are measured independently on both sides of the galaxy. Furthermore the drop to small values of the
running dispersion at the largest major axis distances is significant because 1) it is observed over
several sub-sequences of the halo running dispersion, 2) the {\it running dispersion} curve reaches
small values on both sides of the M87 photometric major axis and 3) the measurements of the M87 halo
PNs velocities come from two independent data-sets. In the North of M87, i.e. for $\mathrm{R > 0}$
in Fig.~\ref{moving_sigma}, the outermost PNs velocities were measured by \citet{doherty09}, while
in the South of M87, i.e.  $\mathrm{R < 0}$ in Fig.~\ref{moving_sigma}, the measured velocities are
from \citet{longobardi15a}. The typical velocity errors are 4.2 $\mathrm{kms^{-1}}$ for the PNs from
\citet{longobardi15a}, and $ \sim 3.0\, \mathrm{kms^{-1}}$ for PNs from \citet{doherty09}.

Binning the PN velocities in Fig.~\ref{moving_sigma} in six elliptical bins with major axis
distances in the range $20\, \rm{kpc} < \rm{R} < 170\, \rm{kpc}$, we compute the velocity dispersion
profile for the M87 halo plus IC PNs. We also compute the halo only velocity dispersions in these
bins, using the robust estimator from \citep{mcneil10} as described in \citet{longobardi15a}. These
values are given in Table~\ref{sigma_values}, and the halo dispersions are plotted in
Fig.~\ref{moving_sigma} (full black circles).

For the M87 halo plus IC PNs' LOSV sample, the velocity dispersion increases from
$\sigma_{\rm{halo+ICL}}\simeq 243.6 \pm 69.5\, \mathrm{kms^{-1}}$ at $\rm{R \simeq 20\, kpc}$ to
$\sigma_{\rm{halo+ICL}}=794.6 \pm 67.8 \, \mathrm{kms^{-1}}$, at $\rm{R=120\, kpc}$. For the M87 halo PNs only, the
velocity dispersion is about constant at a value of $\sigma_{\rm{robust\, halo}}=268.9 \pm 48.4\,
\rm{kms^{-1}}$ between $20$ and $70$ kpc.  It then increases to $\sigma_{\rm{robust\, halo}}=361.7
\pm 26.5\, \rm{kms^{-1}}$, at $\rm{R = 90\, kpc}$, and then declines steeply at larger radii,
reaching $\sigma_{\rm{robust\, halo}}=154.6 \pm 36.4\, \rm{kms^{-1}}$ at $R = 170\, \rm{kpc}$.

In Fig.~\ref{moving_sigma}, we also plot the M87 velocity dispersion measurements from the
integrated stellar light using the IFU VIRUS-P \citep{murphy11,murphy14}.  These measurements
indicate a steep increase in the two outermost bins at $\rm{R > 30\, kpc}$. The comparison
  with the PN velocity dispersion suggests that the reason for the rise of the velocity dispersion
  in the IFU kinematics is the contribution of the ICL at large distances where we
  expect the M87 stellar halo surface brightness to decrease rapidly and the ICL to become
  significant.

Having realized the contribution of the Virgo IC stars to the
kinematics of the outer regions of M87, we now focus on the galaxy
halo.  In the following sections we investigate whether the strong
radial dependence observed in the $\sigma_{\rm{robust\, halo}}$ is an
intrinsic property of M87's halo or whether it signals the presence of
additional velocity components.

\subsection{A shell in a sea of stars: the kinematic footprint of the crown of M87}

Direct evidence of a low mass satellite accretion onto the M87 halo comes from the cold features
observed in the projected phase-space of discrete PNs and GCs \citep{longobardi15b,romanowsky12}, as
well as from the orbital properties of GCs \citep[e.g][]{agnello14} and ultra compact dwarfs
\citep{zhang15}.  Once these cold substructures are identified in phase space, one can recover the
kinematics of the main halo component.

In their recent study, \citet{longobardi15b} used Gaussian Mixture Models to statistically
  separate the PNs of their newly discovered \textit{crown} substructure from the LOSVD of the 253
  M87 halo PNs as classified by \citet{longobardi15a}. This resulted in a total of 53 PNs that had
a small average probability ($\gamma_{i}\sim0.3$) to be part of the M87 main halo. We can now use
this information and compute the velocity dispersion of the remaining M87 halo. As seen in
Fig.~\ref{moving_sigma} (red and black full dots, respectively), the sigma profile without the
\textit{crown} does not change substantially. The contribution by the accreted satellite reduces the
LOS velocity dispersion in those radial bins where the number of \textit{crown} stars is largest. It
is clear that the strong radial variation of the velocity dispersion profile remains an intrinsic
property of the M87 main halo.

\begin{table}
\centering

   % \centering
  % \caption{\small{Flames Configuration and Exposure Time}}
   \centering
   
     \begin{tabular}[h]{ c  c  c  c}
     \hline
     \hline
     R & $\sigma_{\rm halo+ICL}$ & $\sigma_{\rm robust\, halo}$ & $\sigma_{\rm halo\,{no\, crown}}$ \\
     &  &  &  \\
     (kpc)& ($\rm{kms^{-1}}$) & ($\rm{kms^{-1}}$)& ($\rm{kms^{-1}}$)\\
     \hline
     20 & 243.6$\pm$ 69.5 &256.7$\pm$33.1  & 269.1$\pm$34.7                  \\
     45 & 358.5$\pm$ 70.0 &301.7$\pm$23.3  & 284.0$\pm$22.7              \\
     70 & 506.4$\pm$ 57.5 &248.3$\pm$26.5  & 290.6$\pm$32.1                 \\
     90 & 691.8$\pm$ 61.0 &361.7$\pm$26.5  & 362.6$\pm$30.4   \\
     120 & 794.6$\pm$ 67.8 &328.1$\pm$37.1  & 254.6$\pm$36.7   \\
     170& -- &154.6$\pm$36.4  & 126.9$\pm$36.6   \\
    
     \hline
        
   \end{tabular}
 
   \caption[\small{Velocity dispersion estimates from the Planetary Nebula sample}]{\small{Velocity
       dispersion estimates from the PN sample in the outer region of M87. Column 1: Major axis
       distance. Columns 2, 3 \& 4: Velocity dispersions and their uncertainties for the M87 halo
       and IC PNs, the M87 halo PNs from the robust estimate of \citet{longobardi15a}, and the M87
       halo PNs with the crown PNs statistically subtracted off (see text for more details)}.}
   \label{sigma_values}
\end{table}

\subsection{Outliers in the M87 halo}
\label{outliers}

In Fig.~\ref{moving_sigma}, we see two pairs of PNs in the southern region of M87 (at negative
distances), where the two PNs in each pair have very similar positions and velocities (red
circles). Both PN pairs are clearly outside the velocity distribution of their neighbours.  We now
determine how likely such velocity configurations are by using conditional probability theory, which
states that the probability of event V$_{i}$ and event V$_{j}$ is
\begin{equation}
\mathrm{P(V_{i}\, and\; V_{j})=P(V_{i})\times  P(V_{j}|V_{i})},
\label{c_probability}
\end{equation}
i.e., the probability of event V$_{i}$ times the probability of event V$_{j}$ given that event
V$_{i}$ occurred. In our case, we can assume that the first PN of each pair is at a random
position and velocity just like most other PNs, so the relevant probability is $P(V_{j}|V_{i})$.

$P(V_{j}|V_{i})$ has two parts, a photometric part $\rm{P_{ phot}}$ that is the probability of finding
a second PN within the measured relative distance $d\rm{D}$ to the first, and a kinematic part $\rm{P_{
  kin}}$ that is the probability of finding it within the measured $d{\rm V}= \|{\rm V}_i - {\rm
  V}_j\|$ from the first PN. The photometric part can be estimated from the PN number density at the
position of the PNe which gives the expected probability of finding one PN within $\pi (d\rm{D})^2$. The
kinematic probability is given by the integral of the normalized LOSVD over the range ${\rm V}_i
\pm d{\rm V}$.  Using the halo PN number density profile from Section~\ref{sec:PNLF} below and a
Gaussian halo velocity distribution centred on $\mathrm{V_{sys}}=1307$ kms$^{-1}$ and with
dispersion $\sigma\sim300\, \rm{kms^{-1}}$, the probability $\rm{P(V_{j}|V_{i})=P_{\rm phot}\times P_{\rm
  kin}}$ of observing both pairs of PNs is $< 0.3 \%$.

These low values support the classification of these PNs as kinematic outliers, even if
their velocities overlap with the range of velocities for the M87 main halo. It is interesting to
notice that these PN pairs close to M87 overlap spatially with a photometric stream in the southern
part of M87, recently identified by \citet{mihos17}. These authors interpreted this photometric
feature as debris from a tidally dissolved dwarf galaxy (marked as small arrow in the central
panel of their Fig.5). Thus from here on the four PNs are flagged as outliers and assigned to the
ICL.

\subsection{Residual contributions from the ICL to the M87 halo LOSVD}
\label{residualICL}
%\subsubsection{The velocity histograms} 
We now analyse in more detail the LOSVD of the remaining M87 halo and investigate whether it
contains residual contributions from the ICL. In what follows, we examine the generalised histogram
where each PN is represented by a Gaussian distribution\footnote{The kernel size is 80
  $\mathrm{kms^{-1}}$, corresponding approximately to $\sigma_{\mathrm{M87}}/4$ , where
  $\sigma_{\mathrm{M87}}=298.4\, \mathrm{kms^{-1}}$ is the velocity dispersion associated to the M87
  halo from the robust procedure in \citet{longobardi15a}. This kernel size represents a compromise
  between faithful structure representation and noise smoothing.}, weighted by its membership
probability, $\gamma_{i}$, to belong to the M87 halo.
   
\begin{figure}
\centering 
\includegraphics[width=8.5cm ]{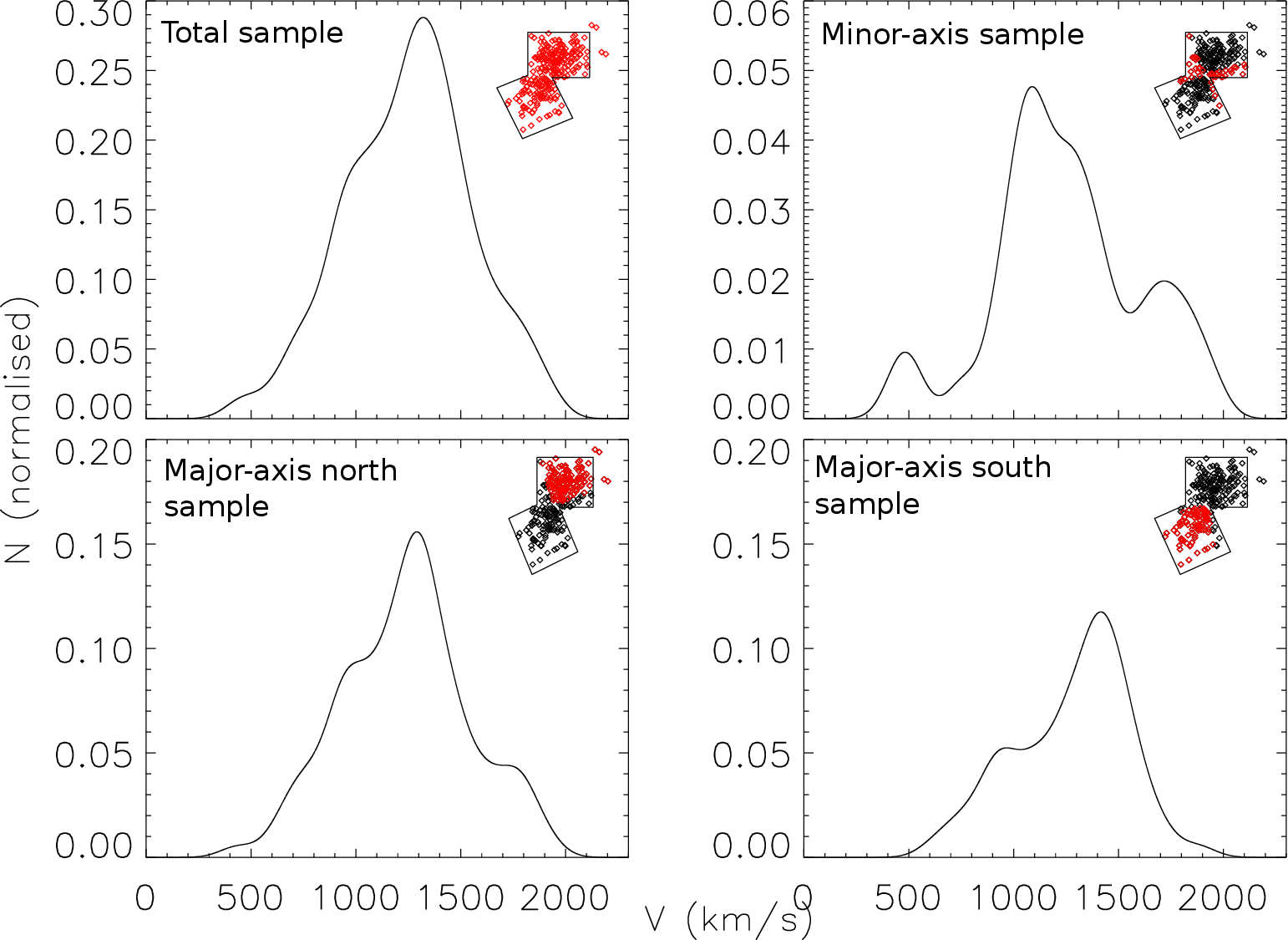}
\caption{\small{Generalised histogram for i) the entire M87 halo PN LOSVD
    (top-left), ii) the subsample along the minor axis (top-right),
    iii) the subsample along the major axis north (bottom-left) and
    iv) south (bottom-right) of M87. The PNs for each subsample are
    depicted in red in the spatial distributions shown in the top
    right corner of each panel. All histograms show a multi peaked
    distribution, with a secondary peak at $\sim$~1000
    $\mathrm{kms^{-1}}$. The probability that such a secondary peak is
    caused by low number statistics is $\sim18\%$ (see text for more
    details).}}
 \label{generalised_hist}
   \end{figure}

In Figure~\ref{generalised_hist} we show the generalised histogram for
four M87 halo PN (sub)samples. These are i) the M87 halo PNs from
\citet{longobardi15a}, ii) the PNs along the minor axis only, and the
PNs along the major axis iii) north and iv) south of the M87
center. See the insets in each panel with the selected PNs depicted in
red. All LOSVDs for the four subsamples show a multi-peaked distribution, with
a second peak observed in all four histograms, centred at $ \mathrm{v_{II}\simeq 1000\, kms^{-1}}$. 

In order to verify whether this second peak is statistically significant we performed a Monte Carlo
analysis.  We simulated 100 LOSVD that were drawn from a single Gaussian distribution with sample
size matching the number of the M87 halo PN sample. For each of the 100 simulated LOSVD, we randomly
extracted three subsamples to simulate the LOSVD along the minor axis, the major axis north, and
south. Only 18\% of the simulated sets of LOSVDs show a double peaked structure in all four
subsamples. Hence, we are confident at the 82\% level ($1.5\times \sigma$) that the feature
correspondent to the second peak is real.

In Fig.~\ref{generalised_hist} the distributions along the minor axis
and major axis north show an additional third peak at
$\mathrm{v_{III}\simeq1800\, \mathrm{kms^{-1}}}$ . However, we note
that the latter feature is more likely to occur as the result of low
number statistics, with a probability higher than 50\%, i.e. less
than $1.0\times \sigma$.

   \subsubsection{A Gaussian Mixture Model for the M87 halo LOSVD:
     Identification of the ${\rm v_{II}}\simeq 900$ $\mathrm{kms^{-1}}$
     peak }\label{sec:GMM}
   
Following \citet{longobardi15b}, we assume that the remaining  M87 halo LOSVD can
be described by a mixture of $K$ Gaussian distributions, and use a
Gaussian Mixture Model (GMM) to identify the individual kinematic
structures (see \citet{pedregosa11}, \citet{longobardi15b} for more
details). GMM implements the expectation-maximisation (EM) algorithm
for fitting mixture-of-Gaussian models. However, our estimated
Gaussian mixture distribution starts from a weighted sample, as we
already removed the {\it crown} contribution statistically.  We 
modified the GMM routine accordingly. 

In this case we have a set of weighted data, $x_i$, the
PN velocities in our case, where each measurement has a corresponding weight,
$\gamma_{i}$. We would like to estimate the parameters of a
Gaussian mixture distribution using this set of weighted data. The
Gaussian density function (PDF) can be written as:

\begin{displaymath}
p(x) = \sum_{k=1}^{K}{{p_{k}(x\, |\, \mu_{k},\sigma_{k})P_{k}}}
\end{displaymath}
where $p_{k}(x\, |\, \mu_{k},\sigma_{k})P_{k}$ is the individual
mixture component centred on $\mu_{k}$, with a dispersion
$\sigma_{k}$, and $P_{k}$ is the mixture weight. The EM procedure then
becomes:

\begin{enumerate}

\item{{\bf E-step}: Compute the posterior probabilities of the
    $i^{\rm th}$ measurement to belong to the $k^{\rm th}$ Gaussian
    components at step $m$, weighted by $\gamma_i$, 
 
\begin{displaymath}
\Gamma_{i,k}^{m}=\frac{p_{k}(x_i\, |\, \mu_{k},\sigma_{k})P_{k}^{m}}{p(x_i)}\times \gamma_{i}
\end{displaymath}}  

where $p(x_i) = \sum_{k=1}^{K}{{p_{k}(x_{i}\, |\, \mu_{k},\sigma_{k})P_{k}^{m}}}$.\\

\item{{\bf M-step}: Compute the new parameter estimates 

\begin{displaymath}
P_{k}^{m+1} = \frac{\sum_{i=1}^{n}{\Gamma_{i,k}^{m}}}{\sum_{k=1}^{K}{\sum_{i=1}^{n}{\Gamma_{i,k}^{m}}}}
\end{displaymath}

\begin{displaymath}
\mu_{k}^{m+1} = \frac{1}{\sum_{i=1}^{n}{\Gamma_{i,k}^{m}}}\sum_{i=1}^{n}{\Gamma_{i,k}^{m} x_{i}}
\end{displaymath}

\begin{displaymath}
\sigma_{k}^{m+1} = \frac{1}{\sum_{i=1}^{n}{\Gamma_{i,k}^{m}}}\sum_{i=1}^{n}{\Gamma_{i,k}^{m} \left(x_{i}-\mu_{k}^{(m+1)}\right)\left(x_{i}-\mu_{k}^{(m+1)}\right)}
\end{displaymath}}

\end{enumerate}

Hence, the E-step in the EM process has been modified such that that the
posterior probability of the $i^{\rm th}$ measurement to belong to the
$k^{th}$ Gaussian at the step $m$, $\Gamma_{i,k}^{m}$, always carries
the starting weight  $\gamma_{i}$ of that measurement. At the end of the
algorithm, each PN's velocity is allocated a posterior probability, that
quantifies its association to each Gaussian component, denoted
by $\Gamma_{i,k}$. Subsequently we use $\Gamma_{i}=\Gamma_{i,1}$ to denote
the probability of belonging to the smooth halo component.

\begin{figure*}
\centering 
\includegraphics[width=6.5cm]{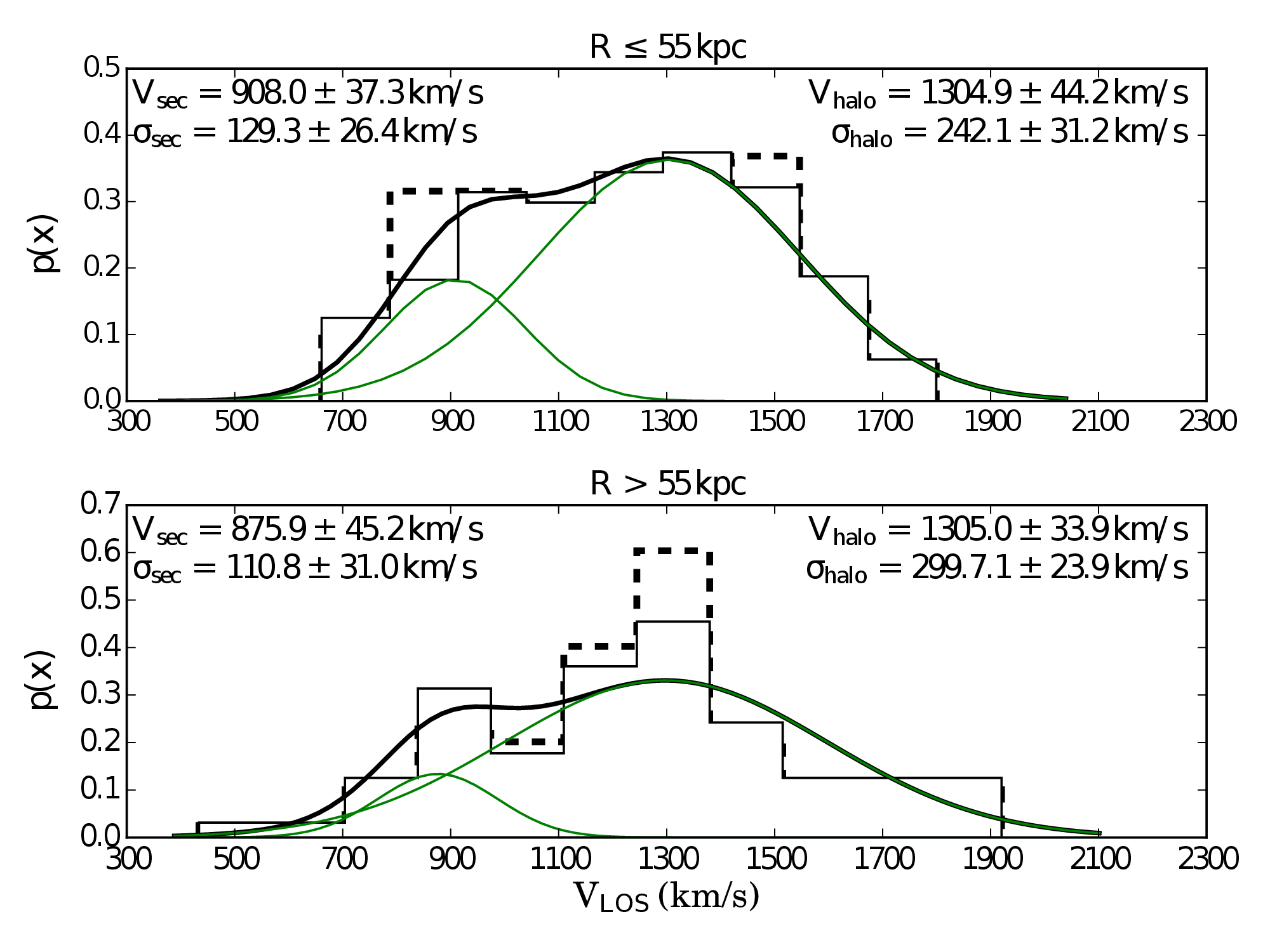}
\includegraphics[width=6.5cm]{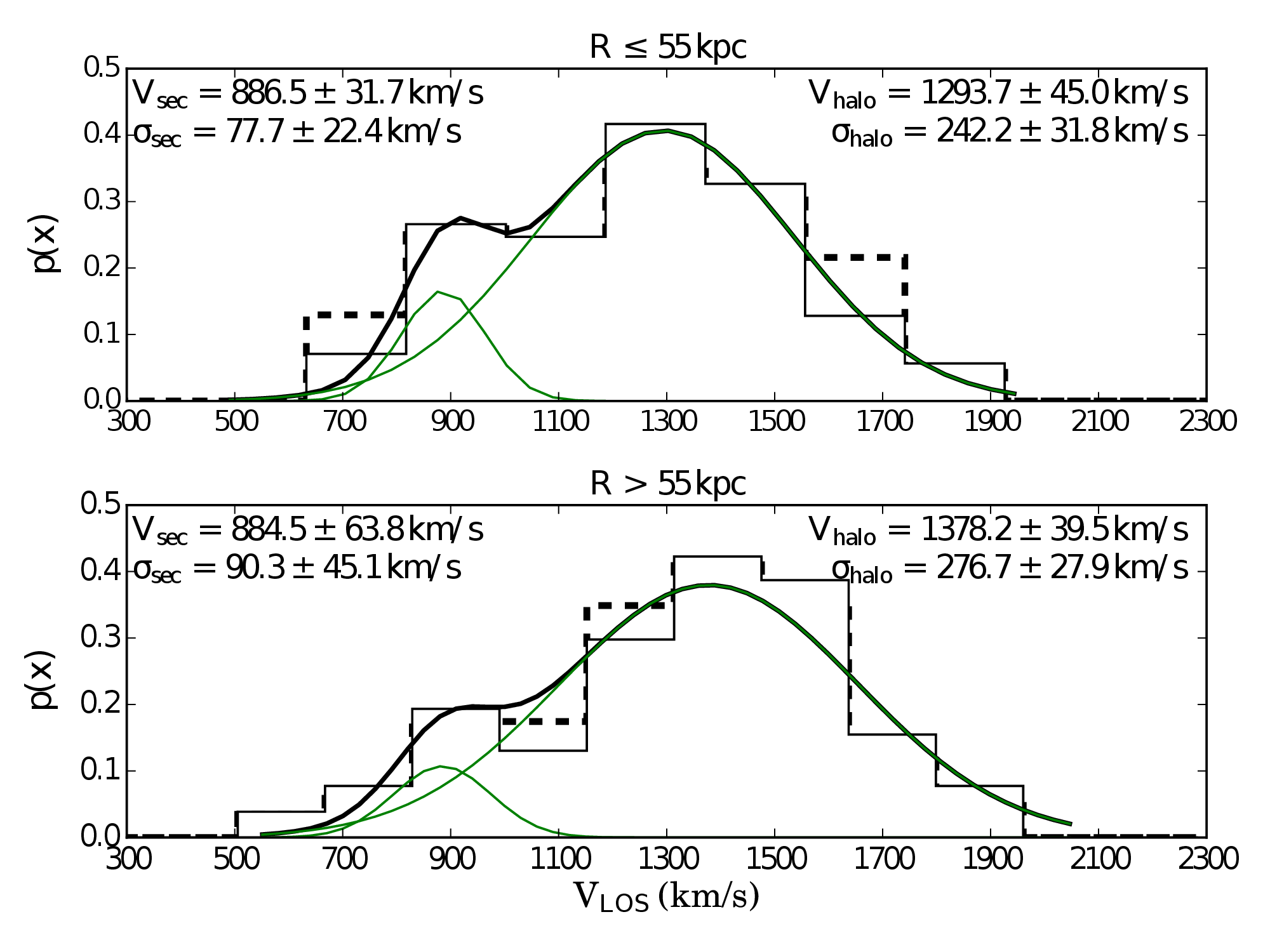}\\
\includegraphics[width=6.5cm]{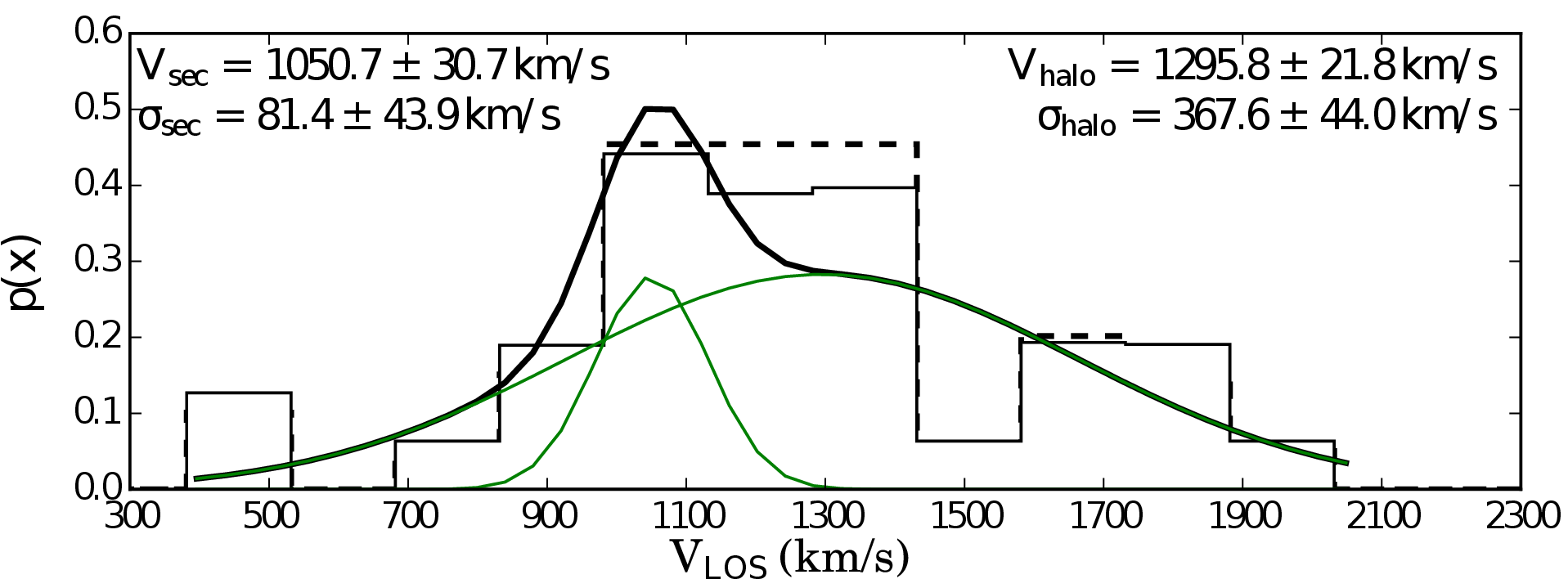}

\caption{\small{Histograms of the LOSVD along the M87 major (top four panels) and minor axis
    (bottom-central panel). The larger number of tracers associated with the M87 halo allows us to
    divide the halo PN sample into northern (top-left panels) and southern subsamples (top-right
    panels), moreover the PN subsamples along the major axis are further divided in two elliptical
    bins as given in the panels. The best-fit GM model (thick black line) identifies in all
    subsamples two Gaussian components (green lines). Dashed histograms represent the PN LOSVD prior
    to the statistical subtraction of the 'crown' substructure.}}
 \label{GMM_decomposition}
\end{figure*} 

We run the GMM on the M87 halo PN subsamples along the minor axis, and the major
axis north and south (Fig.~\ref{generalised_hist} top, right and bottom
panels). Moreover, as we have higher number statistics, the major axis
samples are further divided in two elliptical bins covering radial
ranges $\mathrm{R} \le 55\, \mathrm{kpc}$ and
$\mathrm{R} > 55\, \mathrm{kpc}$, respectively. The GMM identifies the
double peak structure in all these subsamples, for which we show the
histograms of the data along with the best-fit GMM in
Fig.~\ref{GMM_decomposition}.
   
The M87 halo PN LOSVD is then decomposed into two Gaussian mixtures. The main component contributes
about 80\% of the total PN sample and is centred at the systemic velocity of M87, ${\rm v_{sys} =
  1307}$ $\mathrm{kms^{-1}}$, see Table~\ref{GMM_decomposition}. Within the uncertainties, there is
no significant variation of its central velocity along the northern side of the galaxy; however,
south of M87 and for $\rm{R > 55\, kpc}$, it peaks at $\rm{1378.2\pm39.5\, kms^{-1}}$, suggesting
the presence of ordered motion along the LOS at these distances; see Sect.\ref{sec3} for a more
detailed analysis. The velocity dispersion values averaged over the southern and northern major axis
increase from $\rm{\sigma_{halo}=242.1\pm22.3 \, kms^{-1}}$ at $\rm{R\le55\, kpc}$ to
$\rm{\sigma_{halo}=299.7\pm26.2 \, kms^{-1}}$ for $\rm{R\le55\, kpc}$.  Along the minor axis, the velocity
dispersion of the main component is larger, with $\rm{\sigma_{halo}=367.6\pm44.0\, kms^{-1} }$, but
still within 1.5 $\sigma$ of the values measured in the outermost bin along the major axis.

The second Gaussian component is centred at $\rm{v_{II} = 888.7\pm 23.1\, kms^{-1}}$, with a nearly
constant velocity dispersion, $\rm{\sigma_{II}=97.9\pm 15.6\, kms^{-1}}$. In the following, we
  denote this as the $\rm{v_{II}\simeq 900\, kms^{-1}}$ component. Its mean velocity is
constant along the major axis within the uncertainties.  However, along the minor axis it has a
higher value: $\rm{v_{\rm II,\, minor\, axis}=1050.7\pm 30.7\, kms^{-1}}$. The contribution of the
secondary component to the total LOSVD does not vary accross the galaxy and contributes a total of
24 PNs, representing $\sim$10\% of the PN sample associated with the M87 halo in
  \citet{longobardi15a}. For completeness, all the Gaussian fitting parameters are listed in
Table~\ref{GMM_decomposition_table}.

\begin{table}
  
   \centering

     \begin{tabular}[h]{c c c c c}
     \hline
     \hline
     R & $\rm{V_{halo}}$ &$\rm{w_{halo}}$&$\rm{V_{II}}$&$\rm{w_{II}}$\\
     &$\rm{\sigma_{halo}}$ &&$\rm{\sigma_{II}}$ &\\
     (kpc)&$(\rm{kms^{-1}})$&(\%)&$(\rm{kms^{-1}})$&(\%)\\   
     \hline
     &&Major axis north&&\\ 
     \hline   
    $\rm{R \le 55}$&$1304.9\pm44.2$& 89 &$908.0\pm37.2$&11\\
    &$ 242.1\pm31.2$&&$129.3\pm26.4$&\\
     $\rm{R > 55}$&$1305.0\pm33.9$& 87 &$875.9\pm45.2$&13\\
    &$ 299.7\pm23.9$&&$110.8\pm31.0$&\\
     \hline
     &&Major axis south&&\\ 
     \hline   
    $\rm{R \le 55}$&$1293.7\pm45.0$& 88 &$886.5\pm31.7$&12\\
    &$ 242.2\pm31.2$&&$77.7\pm22.4$&\\
     $\rm{R > 55}$&$1378.2\pm39.5$& 90 &$884.5\pm63.8$&10\\
    &$ 276.7\pm27.9$&&$90.3\pm45.1$&\\
     \hline
     &&Minor axis&&\\ 
     \hline   
    &$1295.8\pm21.8$& 90 &$1050.7\pm30.7$&10\\
    &$ 367.6\pm44.0$&&$81.4\pm43.9$&\\
    \hline

\end{tabular}

\caption{\small{Gaussian fitting parameters from the GMM decomposition of the M87 halo PNs' LOSVD
    for the subsamples along the minor axis, major axis north and major axis south. Column 1: Major
    axis distance. Columns 2, 3, 4, and 5: mean velocity, velocity dispersion, and weight of each
    mixture component identified by the GMM. }}
\label{GMM_decomposition_table}  
\end{table}

\subsubsection{The ICPN LOSVD: Comparison with the velocity
  distribution of the galaxies in the Virgo cluster core}
\label{sec:ICLVD}

We now ask the question about the origin of the second Gaussian component in the M87 halo LOSVD.
Dynamical studies of the bright central regions of non rotating elliptical galaxies show that their
LOSVDs are nearly Gaussian, with deviations of the order of 2\% \citep{gerhard93,bender94}. The
secondary component in the M87 halo LOSVD contributes 10\% of the total: it clearly represents a
larger deviation from a single Gaussian velocity distribution!  We note that its average velocity,
${\rm v_{II} = 888.7 \pm 23.1\, kms^{-1}}$, is close to the mean value determined for the velocity
distribution of galaxies in the Virgo subcluster A \citep{binggeli87}. This suggests that this
kinematic component could be a part of the Virgo ICL.

\begin{figure}[!h]

\centering 

\includegraphics[width=9.cm ]{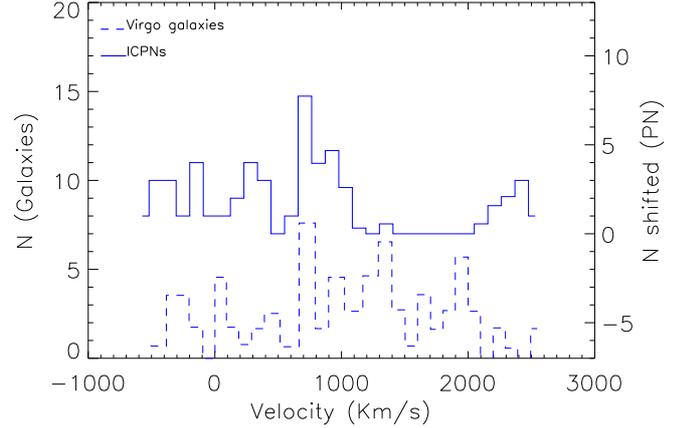}
\caption{\small{Comparison between the combined ICPN LOSVD constructed in Section~\ref{sec2}
    (continuous blue line) and the LOSVD of the galaxies in the Virgo subcluster A region within 2
    deg of M87 (dashed blue line; \citealt{binggeli85, binggeli87}). The ICPN histogram is
      shifted up by $\Delta$N=7 for clarity. The lack of ICPN velocites in the M87 halo range
      of velocities $1100\, \mathrm{kms^{-1}} < \mathrm{V_{LOS}} < 2000\, \mathrm{kms^{-1}}$ is
    related to the lack of information in this range to statistically assign PNs to the ICL
    component (see Section~\ref{sec:ICLVD}). A K-S test in the velocity range $\mathrm{V_{LOS}
      \le 1100\, \mathrm{kms^{-1}}}$ returns a high probability, 97\%, that the ICPN LOSVD is drawn
    from the same distribution as that of the Virgo galaxies.}}
 \label{KS_test} 
   \end{figure}

We can sharpen the argument further by comparing the LOSVD of all identified ICPNs, including the
${\rm v_{II}\simeq 900\, kms^{-1}}$ component, to the LOSVD of the galaxies in the Virgo subcluster A
around M87. Because the ICPNs are not yet dynamically relaxed, we expect that their LOSVD may still
resemble that of the galaxies from which they likely originate.  For the comparison we compute the
LOSVD of all galaxies in the Virgo subcluster A \citep{binggeli85, binggeli87} within $\sim2$ deg of
M87. Figure~\ref{KS_test} shows that these galaxies have a distinctly non-Gaussian LOSVD, with
multiple narrow peaks and broad asymmetric wings, broadly similar to the ICPN. See also the
discussion in \citet{doherty09}.

To carry out a quantitative assessment, we use a Kolmogorov-Smirnov (K-S) test between the LOSVDs of
the Virgo subcluster A galaxies and the ICPNs. The latter includes the 45 PN velocities in the
broad asymmetric wings identified by \citet{longobardi15a}, the 24 PN velocities from the ${\rm
  v_{II}\simeq 900\, kms^{-1}}$ component identified by the GMM analysis in Sect.~\ref{sec:GMM}, and
the 2 pairs of high velocity PNs identified as kinematic outliers in Sect.~\ref{outliers} .  We
carry out the K-S test in the velocity range $\mathrm{v_{LOS} \le 1100\, kms^{-1}}$ because for
$1100\, \mathrm{kms^{-1}} < \mathrm{v_{LOS} < 2000\, \mathrm{kms^{-1}}}$ the GMM did not have enough
information to identify ICPNs that overlap there with most of the M87 halo PNs.  The result of the
K-S test gives a 97\% probability that the ICPN LOSVD is drawn from the same underlying distribution
as that of the LOSVD of the Virgo galaxies around M87. The comparison of the two LOSVDs is shown in
Fig.~\ref{KS_test}.

The asymmetry and skewness of the LOSVD of the galaxies in the Virgo
core and ICL could arise from the merging of subclusters along the LOS
as described by \citet{schind93}.  In their simulations of two merging
clusters of unequal mass, the LOSVD is found to be highly asymmetric
with a long tail on one side and a cut-off on the other side, shortly
($\sim10^9$yr) before the subclusters merge. Around M87, the long tail
is towards small and negative LOSVs, and the cut-off is at positive
velocities, consistent with the merging of the two subclusters centred
around M87 and M86 \citep{doherty09}.

\subsubsection{The PNLF and spatial density profiles of the ICL and the M87 halo}
\label{sec:PNLF}

In this section, we describe the effect of the reclassification of the $\rm{v_{II}\simeq 900 \,
  kms^{-1}}$ component as ICL on the PN luminosity function (PNLF) and on the number density
distributions of the M87 halo PNs and ICPNs.

\cite{longobardi15a} showed that their kinematically separated halo PN and IC PN populations
  had different PNLFs.  The IC PNLF differed from the M87 halo PNLF by having a small value of the
$c_2$ parameter in the generalized PNLF formula \citep{longobardi13} and by the presence of a
morphological signature denoted as ``dip'', located about $1-1.5$ magnitudes below the bright cut-off
of the PNLF.  PN population studies relate the presence of this ``dip'' to recent star formation
\citep{jacoby02,ciardullo04,hernandez09,reid10,ciardullo10} .  The recent work of
  \citet{gesicki18} showed that this dip appears in the PNLF for predominantly opaque nebulae in
  intermediate stages of expansion.

\begin{figure}
  \centering
  \includegraphics[width=8.5cm]{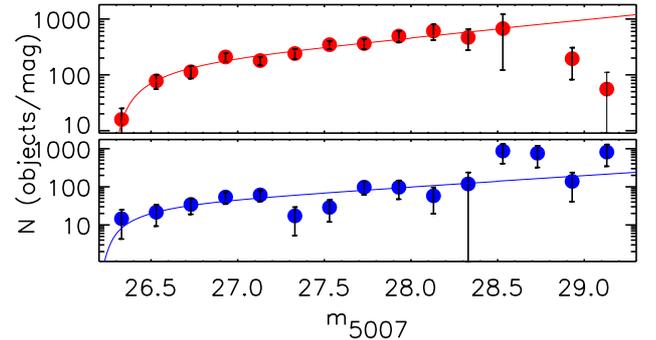}
  \vspace{0.7cm}
  \caption{\textbf{Top panel}:~The luminosity function of the M87 halo PNs (red circles) is shown
    together with the fit of the generalised analytic formula to the halo PNLF, with $c_2=0.72$ and
    bright cut-off at magnitude 26.3. After the statistical subtraction of the reassessed ICL
      component no residual of a dip is seen.  \textbf{Lower~panel}:~PNLF for the ICPNs (blue
    circles). The blue line shows the fit of the generalised analytic formula to the ICPNLF, with
    $c_{2}=0.66$ and bright cut-off magnitude at 26.3. The statistical significance of the dip at
    $1-1.5$ mag fainter than the bright cut-off is enhanced compared to \citet{longobardi15a}.}
\label{PNLF}
\end{figure}

In M87 we find that once the $\rm{v_{II}\simeq 900 \, kms^{-1}}$ component is removed, the halo
  PNLF no longer shows any sign of a ``dip'': see Fig.~\ref{PNLF}.  In this figure, the empirical
PNLF is shown together with the fit of the generalized analytic formula for the PNLF, with the
bright cut-off at 26.3 magnitude and $c_2=0.72$. When the 24 PNs from the $\rm{v_{II}\simeq 900 \,
  kms^{-1}}$ component as well as the 4 kinematic outliers (see Sect.~\ref{outliers}) are merged
with the previously identified ICPN sample from \cite{longobardi15a}, the IC PNLF ``dip'' has higher
statistical significance compared to the earlier analysis, shown in the lower panel of
Fig.~\ref{PNLF}. We conclude that the morphology of the PNLFs thus provides independent support to
the classification of the $\rm{v_{II}\simeq 900 \, kms^{-1}}$ component as part of the ICL.

\begin{figure}
  \centering
  \includegraphics[width=8.5cm]{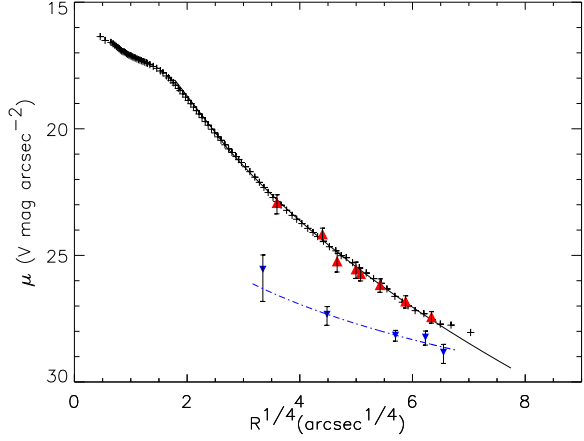}
  \caption{:~Number density profiles for the M87 halo PNs (red triangles) and revised ICPNs (blue
    triangles). The ICPN profile is described by a power-law profile $\mathrm{I_{ICL} \propto
      R^{- \alpha}}$ with $\alpha_{\rm{ICL}} = 0.79 \pm 0.15$ (blue dashed-dotted line).  As in
    \citet{longobardi15a} it is shallower than the halo PN profile, which closely follows the surface
    brightness profile of M87 (black crosses and continuos black line from \citet{kormendy09})
    except in the very outer regions. }
\label{density_profile}
\end{figure}

We conclude this section by comparing the revised number density profiles of the M87 halo and
IC PNs with the surface brightness profiles of M87; see Figure~\ref{density_profile}.  The IC PN
profile now has a slightly steeper gradient than previously quantified by \citet{longobardi13,
    longobardi15a} but remains shallower than that of the M87 halo PNs.  It is fitted by a
power-law $\mathrm{I_{ICL} \propto R^{- \alpha}}$ with $\alpha_{\rm ICL}  = 0.79 \pm 0.15$, so it is not
consistent with a flat distribution.

\subsection{Summary: the line-of-sight velocity distribution for the M87 smooth halo }

On the basis of the robust sigma and the GMM analysis in this Section,
we identified the PNs outliers and PNs associated with either the
revised ICL component or the {\it crown} substructure. Each PN
velocity measurement then comes with a probability of belonging to the
M87 smooth halo. Thus we can determine the first (velocity) and second
(velocity dispersion) moment of the LOSVD in different regions of the
sky and build the corresponding 2D maps. These are the goals of the
next section.

%SECTION 3
\section{Two-dimensional kinematics of the M87 smooth halo: ordered
  vs. random motions }
\label{sec3}

\subsection{Two-dimensional average velocity map}
\label{sec:3.1}

In this section we investigate the average properties of the PN
kinematics of the M87 smooth halo. We build a probability weighted
two-dimensional average velocity field, using an adaptive Gaussian
kernel that matches the spatial resolution to the local density of
measurements \citep{coccato09}, and weights each PN velocity by the
membership probability $\Gamma_{i}$ of the PN to belong to the M87
smooth halo component (see Sect.~\ref{sec2}).

\begin{figure}
\centering

\includegraphics[width=10.cm, clip=true, trim=0cm 2.5cm 0cm 2.5cm]{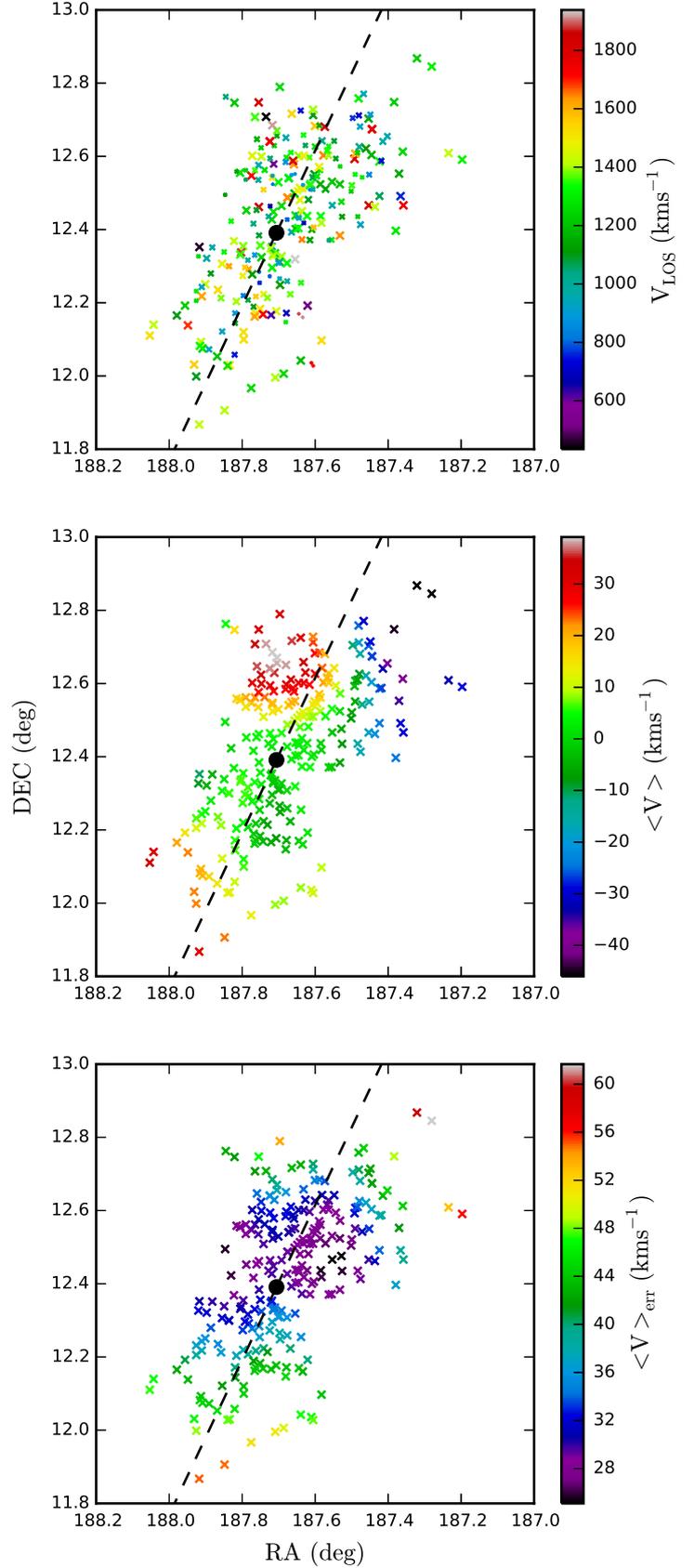}
\caption{\small{\textbf{Top Panel:} Spatial distribution of the 298 spectroscopically confirmed PNs
    in the halo of M87, colour coded according to their $\rm{V_{LOS}}$ and their size scaled
    according to their probability to belong to the smooth halo (top panel). The center of M87
      is shown by the black circle, and the photometric major axis of the galaxy by the dashed
      line.  \textbf{Middle Panel}: Smoothed mean velocity field for M87 using the probability
    weighted Kernel average of the PNs.  \textbf{Bottom Panel}: Errors on the smooth velocity map
    computed by means of Monte Carlo simulations. -- The mean velocity map indicates that the
    kinematics is characterised by ordered motion along the galaxy's major axis. The large positive
    velocities north-east of M87 without counterpart in the south-west are likely due to the
    presence of several PNs with large ($\sim2\sigma$) velocities relative to M87; see text.  North
    is up, East to the left. } }
 \label{velocity_map}
   \end{figure}

At the position of each source ($x_{P},y_{P}$) the mean velocity and
velocity dispersion are:

\begin{equation}
<\mathrm{V} (x_{P},y_{P})>=\frac{\sum_{i}{\mathrm{V_{LOS,i}}w_{P,i}}}{\sum_{i}{w_{P,i}}},
\label{mean_V}
\end{equation}
and
\begin{equation}
<\mathrm{\sigma} (x_{P},y_{P})> =\left[\frac{\sum_{i}{\mathrm{V_{LOS,i}^2}w_{P,i}}}{\sum_{i}{w_{P,i}}}-<\mathrm{V}(x_{P},y_{P})>^2-\Delta V^{2}\right]^{1/2},
\label{mean_sigma}
\end{equation}

\noindent where $\rm{V_{LOS,i}}$ is the $i^{\rm th}$ PN LOSV, and $\Delta V$ is
the instrumental error, given by the median uncertainty on the
velocity measurements, i.e. $\Delta V=4.2\, \rm{kms^{-1}}$; $w_{P,i}$
is the $i^{\rm th}$ PN weight given by:

\begin{equation}
 w_{P,i}=\exp\left({-\frac{D_{i}^2}{2k(x_{P},y_{P})^2}}\times \Gamma_{i}\right),
\end{equation}

\noindent where $D_{i}$ is the distance of the $i^{\rm th}$ PN to $(x_{P}, y_{P})$, and
$k$ is the amplitude of the kernel. Following \citet{coccato09}, $k$
is defined to be dependent on the local density of the tracers, $\rho (x,y)$,
via:

\begin{equation}
k(x,y)=A\sqrt{\frac{M}{\pi \rho\left(x,y\right)}}+B, 
\end{equation}
% What are the ``simulations''
\noindent with $M=20$ representing the number of nearest neighbours
considered in the smoothing technique. $A$ and $B$ are chosen by
processing simulated sets of PNs for a given density, velocity
gradient and velocity dispersion as inferred from the data. The
simulations returned the following values for $A=0.25$ and $B=20.4$
kpc. Thus, each PN is assigned a weight which depends on the
distance, the amplitude of the kernel (in turn depending on the
local tracer's number density\footnote{See \citet{coccato09} for a
  full description of the smoothing technique.}), and on its
probability to belong to the M87 smooth halo component.  As described
in \citet{coccato09,coccato13}, we can also associate errors on the
derived smoothed velocity field by generating 100 different data sets
of mock radial velocities with the same positions on the sky as for the
real sample of PNs. As the same smoothing procedure is applied to the
synthetic data, the statistics of these simulated velocity fields give
us the error associated at the smoothed velocity values at the PN
positions in our field.

In Fig.~\ref{velocity_map}, we plot the positions of the M87 halo PNs on the
sky, colour coded on the basis of their LOSV values. The sizes of their
symbols are proportional to their probability, $\Gamma_{i}$, to belong
to the M87 smooth halo (large symbols; tiny symbols are used instead for
ICPNs). The resulting mean velocity field is given in
the central panel. The galaxy's inner regions are dominated by random
motion, with the mean velocity centred on the systemic velocity of
M87, $1307 \kms$. At large radii the system becomes more complex. There are ordered
motions along the photometric major axis, with approaching velocities to the
north-west, and receding velocities to the south-east side of
M87. Large velocity values are also measured in the north-east
regions, without any symmetric counterpart to the south-west.

From Fig.~\ref{velocity_map} (central panel), it is clear that the
amplitude of the ordered motions along the major axis is small, of the
order of $\sim 30\, \rm{km s^{-1}}$. Such an amplitude is within the
level of uncertainties, as shown by the error map in the bottom panel
of Fig.~\ref{velocity_map}. However, because of its symmetric
properties, we consider it real, and it is further analysed in the
next Section.

The smooth velocity values $\sim 30\kms$ obtained to the north-east of
M87 are also consistent with zero, given the uncertainties. As they
appear only on one side of the galaxy, this suggests that here we are
measuring a local velocity perturbation driven by the presence of a
few high velocity PNs at $\sim1800\kms$ (about $2\sigma$ from the
systemic velocity of M87).

\subsection{Does the M87 outer halo rotate?}
\label{sec:3.2}
\begin{figure*}[!h]
   \includegraphics[width=9cm,clip=true, trim=0.5cm 1.cm 0.5cm 1.5cm]{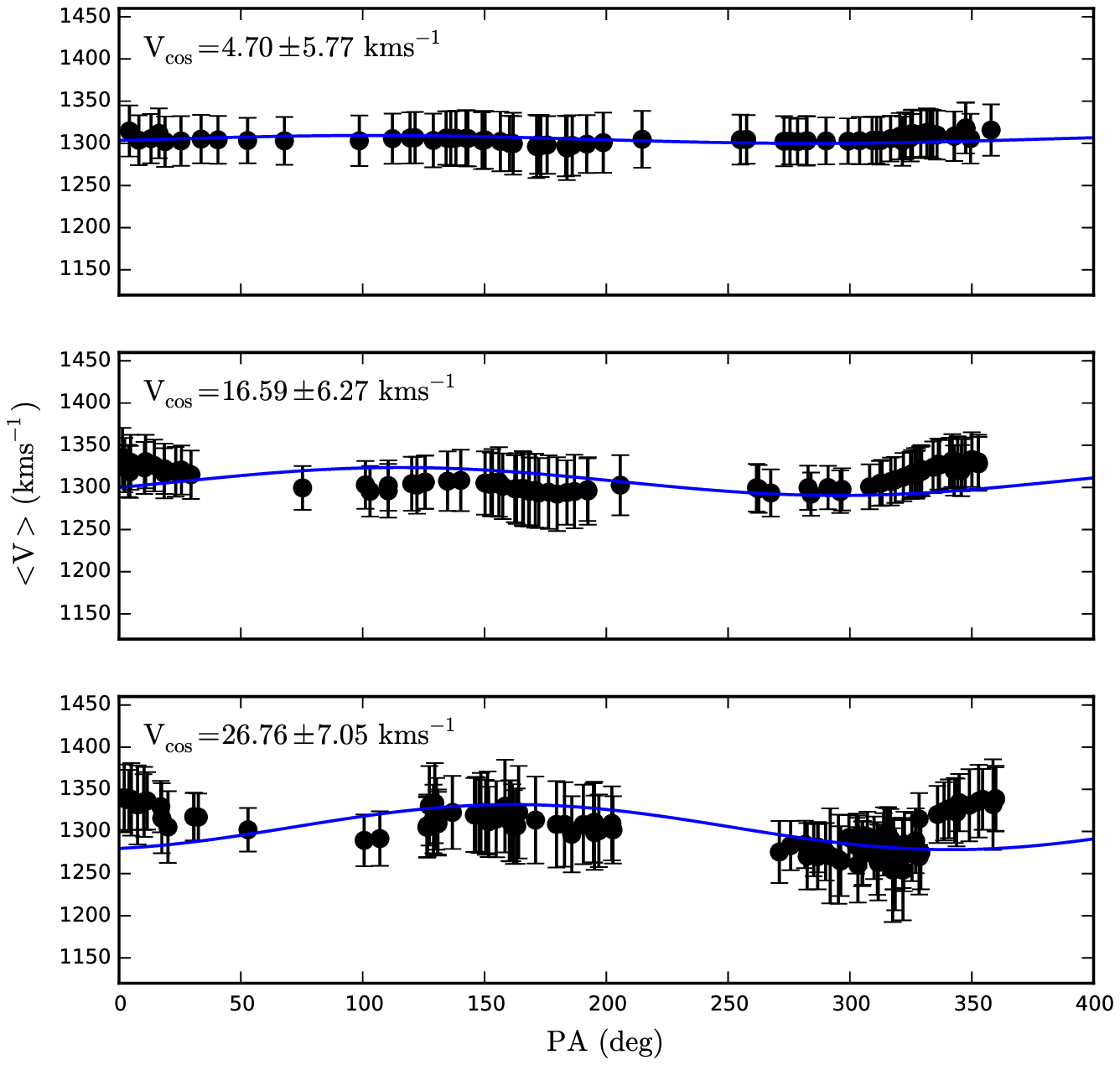}
   \includegraphics[width=9cm,clip=true, trim=0.5cm 1.cm 0.5cm 0.5cm]{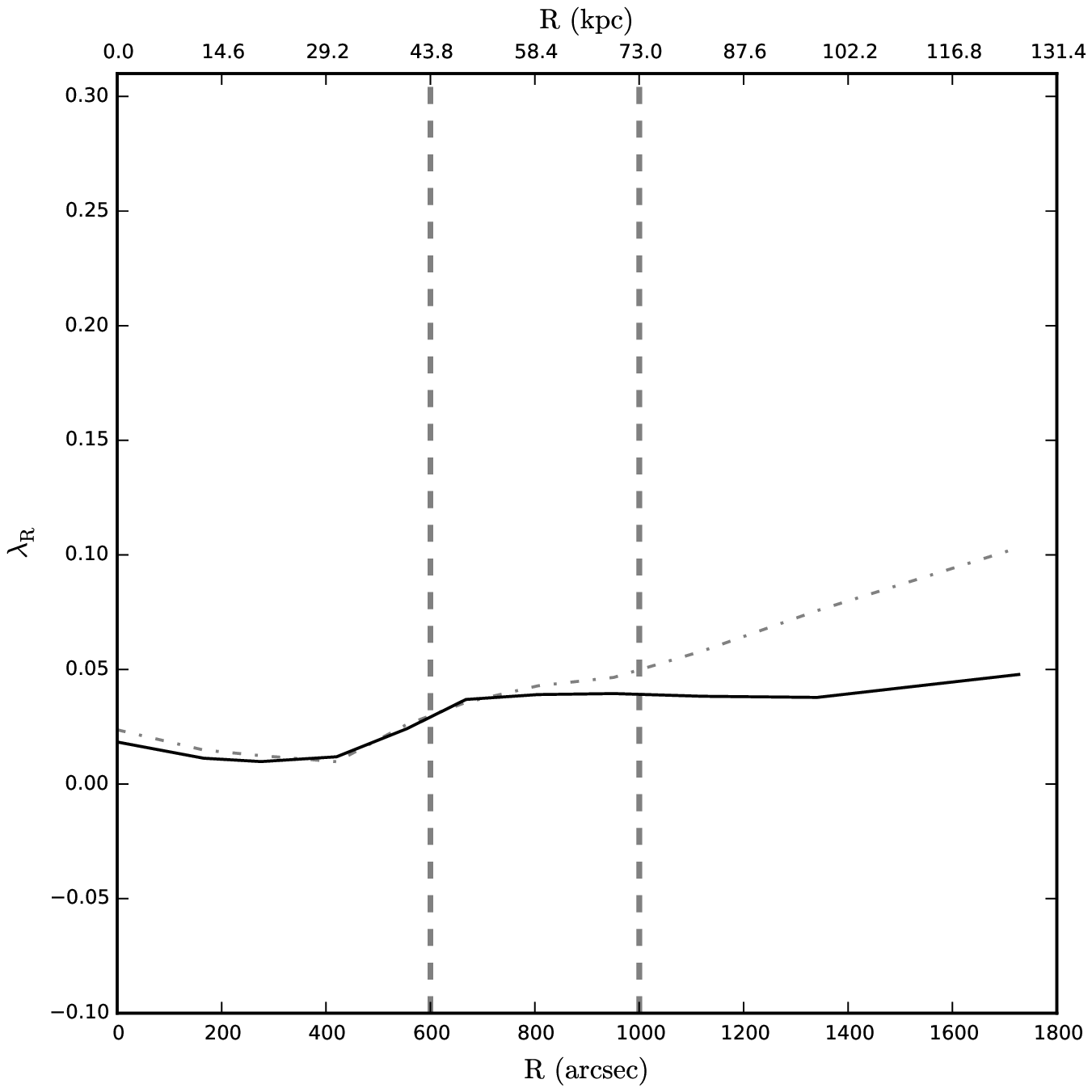}
   \caption{\small{{\bf Left panel:} Mean LOS velocities as a function
       of the PA on the sky for three different elliptical bins (from
       top to bottom we cover major axis distances in the range
       $\rm{R} \le 43.8$ kpc, $43.8 < \rm{R} \le 73.0$ kpc, and
       $\rm{R} \ge 73.0$ kpc, respectively). The continuous blue line
       shows the best-fit model to the data (Eq.\ref{Vfit}),
       consistent with a system that shows increasing importance of
       ordered motion with increasing distance, as shown in the
       plot. The uncertainties on the mean velocities are computed by
       means of Monte Carlo simulations. {\bf Right panel:} Cumulative
       radial $\lambda_{\mathrm{R}}$ profile for the halo of M87
       extracted from PN kinematics. The vertical dotted lines
       identify the major axis distance bins used in the left
       panel. The black continuous line shows the
       $\lambda(\mathrm{R})$ profile computed when approximating the
       mean velocity field with the best-fit cosine model (blue lines
       in the left panels), while the gray dashed-dotted line shows
       $\lambda(\mathrm{R})$ computed from the smoothed velocity data
       points.  At larger radii its amplitude is significantly higher
       than that of the best-fit cosine model, due to the fluctuations
       of the mean velocity values around the systemic velocity (see
       text for more details).} }
       \label{velocity_fit_bin}
 \end{figure*}

The amplitude and axis of rotation are evaluated by approximating the
mean velocity field with that of an axisymmetric rotator. In that case the mean
velocities are modeled by a cosine function of the form:
\begin{equation}
\mathrm{<v_{fit}>(\mathrm{PA},R)=v_{sys}(R)+v_{cos}(R)cos[\mathrm{PA}-PA_{kin}(R)]},
\label{Vfit}
\end{equation}
where R is the major axis distance of each PN from the galaxy's
centre, PA its position angle on the sky \citep{cohen97}. The fitted
values $v_{\mathrm{sys}}$, $v_{\mathrm{cos}}$, and PA$_{\mathrm{kin}}$
represent the M87 systemic velocity, the amplitude of the ordered
motion, and the kinematic PA, with errors derived from fit
uncertainties. To identify possible kinematic decoupling, we divide
our PN sample into three elliptical bins: $\rm{R} \le 43.8$ kpc, $43.8
< \rm{R} \le 73.0$ kpc, and $\rm{R} \ge 73.0$ kpc, and Eq.~\ref{Vfit}
is the fit in each elliptical bin separately.  As shown in
Fig~\ref{velocity_fit_bin}, left panel and Table~\ref{V_cos_table},
the fitted systemic velocities have values consistent with
$v_{\mathrm{sys}}=1307.0\pm7\, \rm{kms^{-1}}$ \citep{allison14}, with
no dependence of the systemic velocity on major axis
distance. Instead, the cosine term increases: for $\rm{R} > 73\,
\rm{kpc}$, the cosine component is $\rm{v_{cos}}=26.76\pm7.05\,
\rm{kms^{-1}}$, and $\rm{PA_{kin}}=161.43^{\circ}\pm
10.64^{\circ}$. With the photometric major axis at
$\rm{PA_{\rm{phot}}}\simeq154.4^{\circ}$
\citep{kormendy09}\footnote{PA are measured with respect the north
  axis, with East to the left.}, the kinematic and photometric axes
are aligned to within the errors.

\begin{table}
\center
\begin{tabular}{c c c c}
\hline
\hline
 \rm{R} &  $v_{\rm{sys}}$ & $v_{\rm{cos}}$ & PA$_{\rm{kin}}$ \\
(kpc) &( kms$^{-1}$)&(kms$^{-1}$)& $(\circ)$\\
\hline\\

$\rm{R} \le 43.8$ &    $1304.62\pm3.81$ & $4.70\pm5.77$ & $104.71\pm70.92$ \\
$43.8 < \rm{R} \le 73.0$&   $1306.99\pm3.87$ & $16.59\pm6.27$ & $115.21\pm18.51$ \\
$\rm{R} \ge 73.0$ & $1305.14\pm4.06$ & $26.76\pm7.05$ & $161.43\pm10.64$ \\
\hline
\end{tabular}
\caption{\small{Cosine model fit parameters in three elliptical
    annuli with increasing major axis distance, R, as described by
    Eq.~\ref{Vfit}. The systemic velocity is constant as a function of
    the distance and in good agreement with the value from the
    literature. At large distance the cosine term V$_{\rm{cos}}$
    increases. The kinematic position angle in the outermost bin,
    PA$_{\rm{kin}}$, is at 160$^\circ$. }}\label{V_cos_table}
\end{table}

 \subsubsection{The cumulative specific angular momentum
   $\lambda_{\mathrm{R}}$ of the M87 halo}
 On the basis of the complete 2D velocity information,
 \citet{emsellem07} introduced the $\lambda_{\mathrm{R}}$ parameter as
 a proxy for the projected specific angular momentum of the stars, defined
 as:
\begin{equation}
\lambda_{\mathrm{R}}=\frac{\sum_{i=1}^{N_\mathrm{p}}\mathrm{F}_{i}\mathrm{R}_{i}\mathrm{|<V>-V_{sys}|}}{\sum_{i=1}\mathrm{F}_{i}\mathrm{R}_{i}\sqrt{(<\mathrm{V}_{i}>-\mathrm{V_{sys}})^2+<\sigma_{i}>^2}},
\label{lambda}
\end{equation}
where F$_{i}$ is the flux associated to the $i^{\rm th}$ point, and
$\mathrm{<V>}$, and $\mathrm{<\sigma>}$ are defined in
Eq.~\ref{mean_V} and Eq.~\ref{mean_sigma}. The $\lambda_{\mathrm{R}}$
parameter measures the significance of rotation as a function of the
distance from the galaxy's centre. Galaxies are then classified as
\emph{fast rotators, $\lambda_{\mathrm{R}}\ge0.1$} (systems
  with aligned photometric and kinematic axes and nearly
  axisymmetric, with a rising $\lambda_{\mathrm{R}}$ profile), and
\emph{slow rotators, $\lambda_{\mathrm{R}}<0.1$} (nearly
  round massive galaxies with a significant misalignment between
  photometric and kinematic axes, moderate degree of triaxiality,
  and a flat or decreasing $\lambda_{\mathrm{R}}$ profile).

We use the probability weighted averaged 2D velocity and
velocity dispersion fields to compute the PN $\lambda_{\rm{R}}$
profile in the surveyed area of the M87 halo. In line with
\citet{coccato09}, the weighting factor $\mathrm{F}_{i}$ is replaced
by $1/c_{R}$ when summing over the PNs. Here the spatial completeness factor $c_{R}$ is taken
from \citet{longobardi15a}. As discussed by \citet{coccato09}, this
procedure incorporates the weighting by the local stellar surface
density by computing a number-weighted sum. In Fig.~\ref{lambda}, we
show the resulting $\lambda_{\rm{R}}$ profile for major axis distances in the
range $\rm{10\, kpc \lesssim R \lesssim 140\, kpc}$ for the M87
halo. The profile is almost flat in the inner $\sim 70$
kpc, and then slowly increases to values of $\sim 0.11$, thus touching
the fast-rotators regime (dotted-dashed line in Fig.~\ref{lambda},
right panel). 

To assess the cause for this increase in $\lambda_{\rm R}$, we
compare the observed $\lambda_{\rm{R}}$ parameter from the 2D field
with that computed using the cosine fit from Eq.~\ref{Vfit}, with
$\mathrm{v_{sys},\, v_{cos},\, and\, PA_{kin}}$ given in
Fig.~\ref{velocity_fit_bin} (left panel). As shown in
Fig.~\ref{velocity_fit_bin} (right panel, black continuous line), the
transition from the inner to the outer regions is still signalled by
an increase of the $\lambda_{\rm{R}}$ profile with major axis
distance. However it flattens at a value of $\lambda_{\rm{R}}\sim 0.05$
and never reaches the $0.1$ threshold. This difference is driven by
the fact that the cosine fit does not represent the apparent streaming
velocity in the North of M87 which we argued above comes from a
few high-velocity PNs with velocities $\sim1800 \kms$.
\footnote{We note that this increase of $\lambda_{\rm{R}}$ profile
  from fluctuations in the mean velocity field is different from that
  described in \citep{wu14} which results from fluctuations in the
  velocity distribution around the local mean velocity.}

Previous studies found that slow rotator galaxies typically increased
their rotation from the central regions into their halos, with some
entering the regime $\lambda_R>0.10$
\citep{coccato09,arnold14,pulsoni18}.  In the case of M87 the
$\lambda_{\rm{R}}$ parameter also rises into the halo but only to values
$\lambda_R\simeq 0.05$, remaining safely in the slow rotator regime.

%SECTION 4
\section{Velocity dispersion profile of the stellar tracers in M87}
\label{sec_sigma_R}

PNs are single stars whose velocities are a discrete realization of the LOSVD of the stellar
population in a given region of a galaxy.  PNs are ubiquitous probes of the kinematics of the parent
stars at radii where the surface brightness is too faint to measure absorption line features with
the required S/N ratio.  Hence they are very well suited to complement the stellar-kinematic
measurements in the inner regions.  In this section, we combine measurements of the second moment of
the LOSVD, $\sigma$, from absorption line kinematics in the inner high surface brightness regions
with those from the PN LOSVDs at large radii, obtaining the velocity dispersion profile out to
$\sim170$ kpc along the major axis.  We also discuss velocity dispersion measurements for the GC and
ultra-compact dwarf (UCD) systems in M87 in comparison with the composite $\sigma$ profile of the
stars.

\subsection{The $\sigma$ profile within 20 kpc radius in M87}
\begin{figure*}[!h]
\centering
   \includegraphics[width=11.0cm]{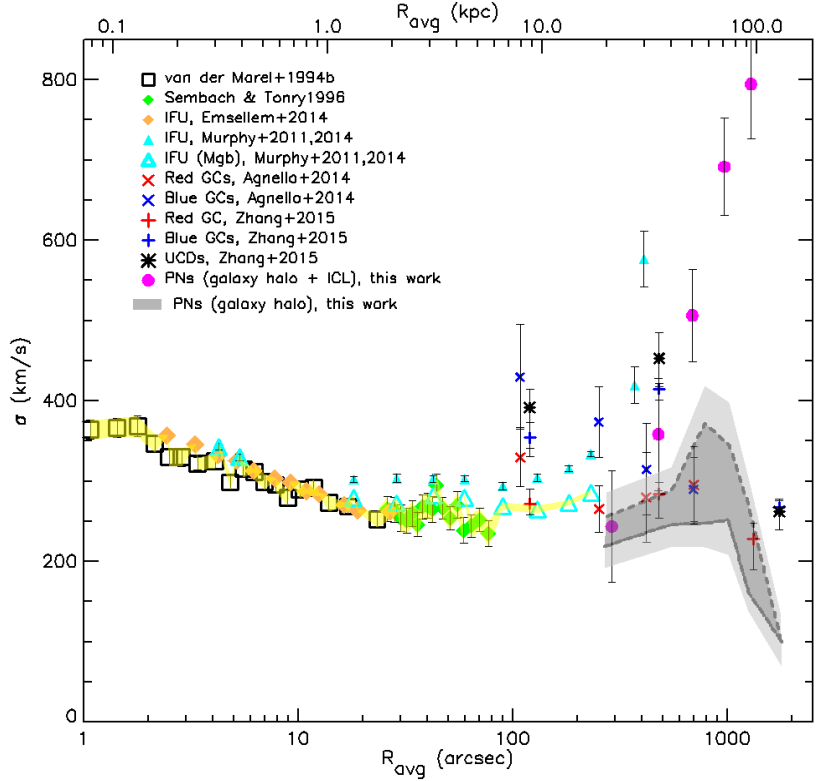}
   \caption{\small{Velocity dispersion profile for the halo of M87 as function of
     the average ellipse radius, $R_{\mathrm{avg}}=(ab)^{1/2}$ of the isophote, (for $\mathrm{R >
       400 \arcsec}$, we have assumed a constant ellipticity of $e=0.4$).  In the inner $80\arcsec =
     5$ kpc, we show the absorption line data from \citet{vandermarel94} (squares), \citet{sembach96}
     (green diamonds), and from \citet{emsellem14} (orange diamonds). Cyan triangles present IFS
     VIRUS-P data from \citet{murphy11,murphy14}, for the case when the velocity dispersion is
     computed making use of the entire spectral region (filled triangles), or when it is calculated
     only from the Mg $b$ region (large open triangles). The yellow-shaded area indicates our
       fiducial velocity dispersion profile of the stars from absorption line spectroscopy (long
       slit/IFS) in the radial range out to $\sim 200'' = 15$ kpc.  The magenta full dots show the
     new PN velocity dispersion values for our whole sample, i.e. without separation between M87
     halo PNs and ICPNs. The fiducial range of $\sigma$ estimates we obtain for the M87 smooth halo
     PNs alone is indicated by the dark gray area, with boundaries given by the robust estimates of
     sigma (black line) and by the simple RMS of the halo PN data (dashed black line). The light
     gray area includes the respective 1-sigma uncertainties for these values added as well.  Thus,
     the PN kinematics trace a $\sigma$ profile for the M87 halo with a strong radial dependence,
     increasing from 20 to 90 kpc and then decreasing strongly, reaching a minimum value of $\sim
     100 \mathrm{ km }^{-1}$ at $\mathrm{R_{av} \simeq 130\, kpc}$ ($\mathrm{R_{\rm maj} \simeq
       170\, kpc}$).  For comparison, red and blue stars show GC velocity dispersion data from
     \citet{agnello14}, while black asterisks, blue, and red crosses indicate sample velocity
       dispersions for UCDs, blue, and red GCs, respectively, as presented in
       \citet{zhang15}}.}  \label{sigma_profile}

\end{figure*}

M87 has been the target of many spectroscopic studies with the goal of determining the integrated
mass profile from the inner regions to the outermost radii. Velocity dispersion measurements from
absorption line spectroscopy (long slits and IFS) available in the literature are reproduced in
Figure~\ref{sigma_profile}.  In this figure, the velocity dispersion profile of the stars in M87 is
plotted as function of the isophotal average ellipse radius, $R_{\mathrm{avg}}=(ab)^{1/2}$, where
$a,b$ are the isophote major and minor axes.

At $R_{\mathrm{avg}}~\leq 5\, ~\rm{kpc}$, the absorption line measurements from
\citet{vandermarel94,sembach96} and \citet{emsellem14} show a characteristic profile for hot stellar
systems.  It has a central peak at $\sigma \sim 400\, \rm{kms^{-1}}$, then declines to a value of
$\sigma \sim 270\, \rm{kms^{-1}}$ at $\sim 2 $ kpc and remains flat out to $\sim10$ kpc. At these radii
the $\sigma$ measurements from the literature agree within their uncertainties; however we note that
the values from \citet{sembach96} were corrected for a $7-10\%$ systematic velocity offset
attributed by the authors to the large slit width adopted for their observations \citep[for more
details see discussion in ][]{sembach96,romanowsky01,doherty09}.

In addition to the MUSE IFS data \citep{emsellem14}, also VIRUS-P IFS data from \citet{murphy11}
are available in this region. At $R > 1$ kpc, the IFS VIRUS-P measurements have a systematic
positive offset of $\sim30\, \mathrm{kms^{-1}}$ with respect to the MUSE and slit data. This offset
is present when the $\sigma$ measurements are obtained from the combined analysis of four wavelength
regions (G-band, H-beta, Mg$b$, Iron; filled cyan triangles in Fig.~\ref{sigma_profile}).  When
$\sigma$ values are measured only in the Mg$b$ region of the spectrum \citep[][open cyan triangles
in Fig.~\ref{sigma_profile}]{murphy11}, then the VIRUS-P and MUSE data sets agree.  Then the IFS
VIRUS-P measurements \citep[open cyan triangles][]{murphy11} extend the $\sigma(R)$ profiles to
larger radii.  In the radial range $5\, \rm{kpc} < \rm{R_{avg}} < 20\, \rm{kpc}$, they signal an
increase of the velocity dispersion values to $\sim 280$ kms$^{-1}$.

At radial distances $10\, \rm{kpc} < \rm{R_{avg}} < 20\, \rm{kpc}$, GCs and UCD
 galaxies have also been identified and their LOSVDs measured. Red and blue GC sample
  velocity dispersions from \citet{agnello14} and \citet{zhang15} are shown as red and blue crosses
  and plus symbols, respectively. While the $\sigma$ values of the red GCs are in better agreement
  with those from the stars (within the uncertainties), blue GC $\sigma$ values deviate
  strongly. The velocity dispersion values for the population of Virgo UCDs in \citep[black
    asterisks;][see Sect.~\ref{discussion} for more discussion]{zhang15} are similar to those
  of the blue GCs except for the outermost point, which is closer to the red GC dispersion.

 \subsection{The velocity dispersion profile of the smooth M87 halo
   from 20 kpc out to 170 kpc}

In Section~\ref{sec2} we investigated the influence of the ICL on the PN LOSVDs at radii $R > 20$
kpc. In the IFS VIRUS-P data, the presence of the ICL is disclosed by the sudden increase of
$\sigma$ from $\sim 300\, \rm{kms^{-1}}$ to nearly $\sim 600\, \rm{kms^{-1}}$
\citep{murphy14} \footnote {In that paper, dispersion values are computed using the information
  coming from the entire spectral region between $4100\AA\ - 5400\AA\ $.}. The comparison of these
values with the {\it running dispersion} of the M87 PN LOSV sample in Fig.~\ref{moving_sigma} shows
very good agreement. We note that, because the ICL is unrelaxed \citep{longobardi15a}, these high
$\sigma$ values include a significant contribution from unmixed orbital motions and do not trace the
enclosed mass only. Therefore, for a proper mass analysis of the M87 halo, the ICL contribution must
be subtracted.
 
In the course of the extensive analysis carried out in Section~\ref{sec2}, we computed the
probability for each PN to be associated with the smooth M87 halo; we can thus use this probability
to compute the histograms of the PN LOSVD for the smooth M87 halo only, in different outer radial
bins. These histograms are shown in Appendix~\ref{apx_histo}. They have limited statistics and may
thus deviate from Gaussian LOSVDs. To characterize the associated uncertainties, we computed the
velocity dispersion values using the robust sigma algorithm\footnote{For a short description of this
  technique see Appendix~\ref{apx_histo}. More details can be found in \citep{longobardi15a}.} and
the direct {\it RMS} values from the PN LOSVDs in these bins. The estimated radii and velocity
dispersions using either method are listed in Table~\ref{sigma_values_smooth_halo} for all bins.
Since for small samples the robust sigma may underestimate the true velocity dispersion due to
overclipping, and the direct {\it RMS} sigma may overestimate it due to its sensitivity to
velocities in the wings of the distribution, we take these two determinations as the boundaries of
our fiducial range of velocity dispersion values from the PN LOSVDs for the smooth M87 halo. This
range is shown by the dark shaded area in Fig.~\ref{sigma_profile}, with the lower boundary from
the robust sigma depicted by the black continuum line and the {\it RMS} estimates by the dashed
line, respectively. The range of velocity dispersions obtained by adding also the $1\times \sigma$
uncertainties of the two determinations is shown as a light gray shaded area.

\begin{table}
\centering
     \begin{tabular}[h]{ c  c  c  c}
     \hline
     \hline
     R$_{\rm robust}$ & $\sigma_{\rm robust}$ & R$_{\rm RMS}$  & $\sigma_{\rm RMS}$ \\
     &  &  &  \\
     (kpc)& ($\rm{kms^{-1}}$) & (kpc)& ($\rm{kms^{-1}}$)\\
     \hline
     25.2 & 218.4$\pm$26.7 & 25.9  & 256.5$\pm$29.3     \\
     51.5 & 245.6$\pm$28.0 & 51.7  & 287.7$\pm$29.3     \\
     74.3 & 248.0$\pm$30.0 & 74.3  & 371.9$\pm$47.3     \\
     95.8 & 252.0$\pm$43.6 & 96.2  & 345.1$\pm$53.0     \\
     118.6& 161.9$\pm$24.5 & 118.4 & 270.4$\pm$34.8     \\
     170.0& 99.6 $\pm$31.5 & 167.2 & 150.4$\pm$31.9     \\
    
     \hline
        
   \end{tabular}
 
   \caption[\small{Velocity dispersion estimates for the smooth M87 halo as function of the major-axis distance}]{\small{Velocity dispersion estimates for the smooth M87 halo as function of the major-axis distance. Column 1\&3: Major-axis distance. Columns 2\& 4: Velocity dispersion and their uncertainties for the M87 smooth halo given by the robust estimate and by the simple RMS of the data (see text for more details)}}.
   \label{sigma_values_smooth_halo}
\end{table}

Our fiducial range of velocity dispersion values for the smooth M87 halo indicates a strong
variation of the $\sigma(R)$ profile with radius. The $\sigma(R)$ profile from PN LOSVDs extend the
slowly rising trend captured by the IFS measurements \citep{murphy14} out to $\mathrm{R_{avg} \simeq
  17\, kpc}$, with $\sigma$ rising to about $280\, \mathrm{kms^{-1}}$ there. The fiducial velocity
dispersion range from PNs indicates a further rise of the dispersion to $\sigma\simeq 300\,
\mathrm{kms^{-1}}$ at $50 < \mathrm{R_{avg}< 70\, kpc}$, followed by a steep decline down to $\sigma
= 100\, \mathrm{kms^{-1}}$, at $\mathrm{R_{avg}\sim 135\, kpc}$ (corresponding major-axis radii are
1.3 times larger).  We note that the rise and steep drop of the PN velocity dispersion profile is
seen on both the NE and SW sides of M87; see Fig.~\ref{moving_sigma}.

In the next section we investigate whether this strong radial variation of the velocity dispersion
profile is consistent with a change in the physical properties of the stellar orbits in these outer
regions in dynamical equilibrium. We approach this problem with an approximate analysis based on the
spherical Jeans equations, connecting the circular velocity curve inferred from X-ray observations
with the surface brightness profile for the smooth M87 halo.

\subsection{The gravitational potential, density, and orbital anisotropy in the outermost halo of M87}
%the comparison with the total enclosed mass from X-ray emissivity maps
\label{Xray}

\def\vrm{\mathrm{v}}
\def\vcirc{{\vrm_c}}
\def\vcopt{{\vrm_{c,\mathrm{opt}}}}

Studies using stellar kinematics, lensing, and X-ray observations \citep[see,
  e.g.,][]{gerhard01,treu06,gavazzi07,churazov10} have indicated that the gravitational potentials
of elliptical galaxies are approximately isothermal. With this assumption, i.e., the circular
velocity $\vcirc$ is constant, \citet{churazov10} showed that the spherical Jeans equation leads to
simple relations between $\vcirc$, the LOS velocity dispersion profile $\sigma(R)$, and the surface
brightness profile $I(R)$. For the case of a system with either isotropic or radial orbital
distribution, the relations between $\vcirc$ and the local properties of $\sigma(R)$ and $I(R)$ are
\begin{equation}
\label{der_eq}
\sigma^{2}_{\mathrm{iso}}(R) = \vrm^{2}_{c}\frac{1}{1+ \alpha + \eta},\\
\sigma^{2}_{\mathrm{rad}}(R) = \vrm^{2}_{c}\frac{1}{(\alpha + \eta)^{2}+\delta-1},
\end{equation} 
where $\alpha$ and $\eta$ are the negative of the logarithmic radial gradients of $I(R)$ and
$\sigma(R)$, and $\delta$ is the second logarithmic derivative of $I(R)\sigma^2(R)$; see eqs.~22, 23
of \citet[][]{churazov10} for more detail.  \citet{churazov10} used the above equations to infer the
circular velocity $\vcopt$ from the optical data $I(R)$ and velocity dispersion profile $\sigma(R)$
at different radii, and to compare with the circular velocity curve $\vrm_{c,X}$ derived from X-ray
emissivity and temperature maps for M87 from \textit{Chandra} and \textit{XMM-Newton}.  Their
analysis led to a best fit relation between these two estimates of $\vcopt\sim 1.10-1.15
\vrm_{c,X}$, implying an average contribution by non-thermal pressure of 20-30\% for six X-ray
bright galaxies, with a particularly large value for M87.

Given our new assessment of the M87 outer halo kinematics, we carried out an independent comparison
of the total enclosed mass profiles from a dynamical estimate using optical data, and from the most
recent X-ray information obtained by combining that data of \citet{churazov10} with those of
\citet{simionescu17} at larger radii. Our approach is the following: we adopt the
circular velocity curve from the X-ray maps out to 2500 arcsec, and predict the LOS $\sigma(R)$
profile from the surface brightness distribution of the {\it smooth and relaxed} stellar halo of M87
using eqs.~\ref{der_eq}.  We derive the expected velocity dispersion curves in the isotropic and
completely radially anisotropy cases, with the goal of comparing with our fiducial $\sigma(R)$ profile
that combines absorption line kinematics with the PN LOS velocity dispersion for the
smooth M87 halo component.

\begin{figure*}[!h]
\centering
   \includegraphics[width=13.cm]{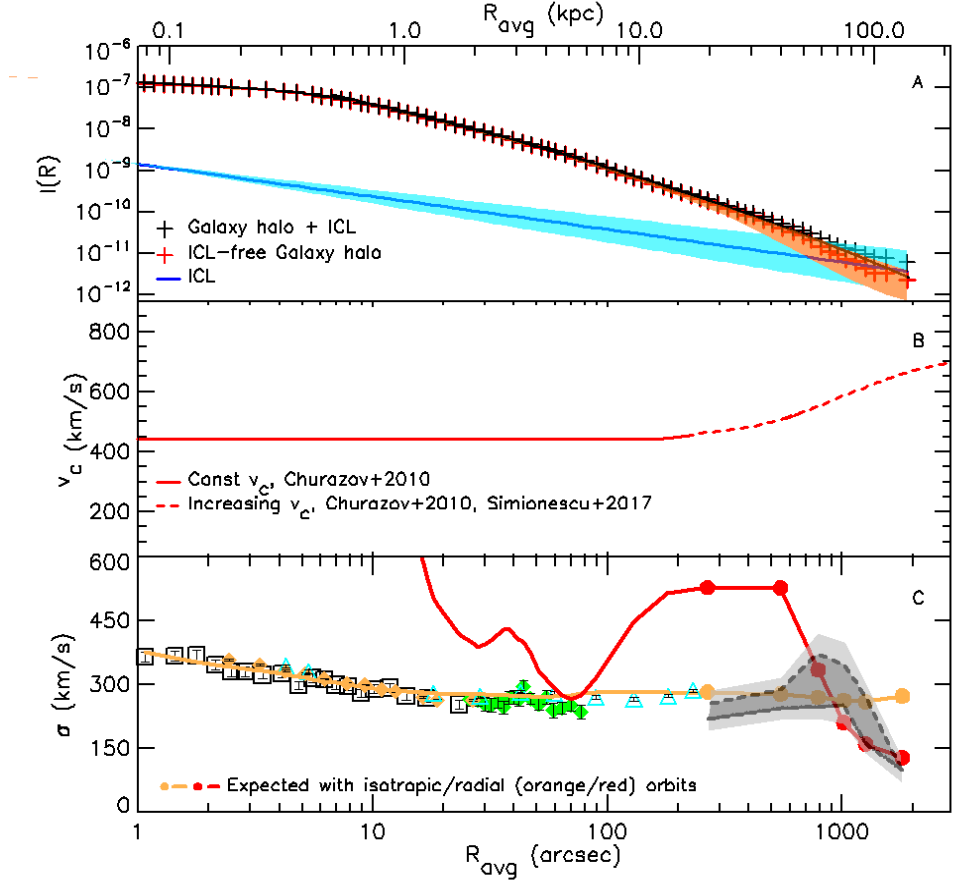}
   \caption{\small{Velocity dispersion profile out to 2000 arcsec compared with predictions
       from Jeans equations.  Panel A shows the surface brightness profile of the M87 smooth halo
       (red crosses and shaded area) obtained from the extended photometry \citep{kormendy09}, after
       subtracting the ICL contribution (blue line and shaded area).  The black continuous line
       shows the Sersic fit with $n=11$ to the M87 halo + ICL extended photometry. The adopted
       circular velocity profile is shown in Panel B. This is a combination of a flat profile (red
       line) out to 200\arcsec and an increasing profile for larger distances (dashed red line, see
       text for more information). Shown in Panel C is the velocity dispersion profile $\sigma(R)$
       from this work (data points, fiducial range from PNs, and legend as in
       Fig.~\ref{sigma_profile}). Large dots represent the expected $\sigma$ values computed from
       eqs.~\ref{der_eq} at the average radii of the radial bins, for isotropic (orange) and
       completely radial (red) orbital anisotropy. The comparison with the fiducial range of
       $\sigma$ values for halo PNs (shaded area) suggests that the distribution of orbits changes
       from near-isotropic at $\sim$200\arcsec to strongly radial at
       $\sim$2000\arcsec.}} \label{circ_v}
 \end{figure*}

Results are shown in Fig.~\ref{circ_v}. In Panel A, we show the surface brightness profile of the
M87 smooth halo, indicated by the red crosses and red shaded area, the latter indicating the
$1\sigma$ uncertainty. This surface brightness profile is computed from the extended photometry of
M87 in \citet{kormendy09} (black crosses, with continuous black line showing the Sersic fit with
$n\sim 11$) by subtracting off the ICL contribution (blue line and light blue shaded area, the
latter indicating its $1\sigma$ uncertainty).  The contribution from the ICL was determined in
Section~\ref{sec:ICLVD} and Fig.~\ref{density_profile} from the kinematical tagging of the ICPNs.
As the inferred ICL surface brightness is of the same order as the surface brightness of the M87
smooth halo at these large distances, it must be taken into account, i.e. subtracted, so as to have
a consistent set of tracer profiles in the Jeans analysis.  We note that this analysis cannot be
done for the combined halo plus ICL, because the LOSVD of the ICL shows that it is not in
equilibrium in the gravitational potential. Because of the extended tails of its LOSVD, it
would lead us to incorrectly infer too large masses at the largest radii.

The adopted $\vrm_{c}=\vrm_{c,X}$ is plotted in Panel B. It is a combination of three parts: (i) a
flat profile out to 200\arcsec ($\sim15$ kpc) fitted to the data from \citet{churazov10}, ii) an
increasing profile obtained after differentiating a smooth non-linear fit to the same M87 X-ray
potential data in the range from 200\arcsec ($\sim15$ kpc) to 1200\arcsec (88 kpc), as presented in
\citet[][their Fig.1]{churazov10}, and iii) from 1200\arcsec onward, the circular velocity
corresponding to the NFW profile fitted by \citet{simionescu17} to their Suzaku data within 400
kpc. The obtained $\vrm_{c}=\vrm_{c,X}$ profile summarizes the observational evidence that the
circular velocity rises more steeply than an isothermal profile at large radii.  We can nonetheless
use the local eqs.~\ref{der_eq} with this circular velocity profile to estimate velocity
dispersions, because $\vrm_{c,X}$ varies only slowly with radius as is confirmed by panel B
considering the logarithmic radius scale.

Panel C then shows the expected velocity dispersions $\sigma_{\mathrm{iso}}$ and
$\sigma_{\mathrm{rad}}$ according to eqs.~\ref{der_eq}, over-plotted on the observed M87 sigma
profile as presented in Fig.~\ref{sigma_profile}.  The comparison between these velocity dispersion
estimates and our fiducial velocity dispersion profile suggests that the distribution of orbits in
the outer halo of M87 changes from near-isotropic at 200 \arcsec ($\sim15$ kpc) to completely radial
at 2300\arcsec ($\sim$ 170 kpc). The radial velocity dispersion curve inferred from eq.~8 has
non-negligible uncertainties because of the errors in the ICL-subtracted brightness profile; thus
further dynamical modelling will be needed to confirm this. Note that the fact that such a dynamical
structure of the M87 halo appears consistent with all data in Fig.~\ref{circ_v} obviates the need
for a truncation of the halo density as inferred by \citet{doherty09}. This is ultimately due to the
better statistics in the new ICPN data which allowed us to subtract the (non-equilibrium) ICL
component from both the surface brightness and LOS velocity dispersion data.

%SECTION 5
 
\section{Discussion} 
\label{discussion}

\subsection{The Virgo ICL. An unrelaxed component in the cluster core}

In Section~\ref{sec2}, we carried out a careful analysis of the PN LOSVs in the velocity range
$500-2000\,\mathrm{kms^{-1}}$ around M87. We identified a velocity component at $\mathrm{v_{II}}\simeq
900\, \mathrm{kms^{-1}}$ and two pairs of outliers as ICPNs. The complete LOSVD for the ICPN
population was then obtained by combining the newly identified $28$ ICPNs with those in the extended
velocity wings from \citet{longobardi15a}.  The resulting ICPN LOSVD around M87 has a peak at
$\mathrm{v_{II}}\simeq 900\, \mathrm{kms^{-1}}$ with extended wings, skewed towards negative
velocities (Fig.~\ref{KS_test}).

The identification of the additional $28$ ICPNs was supported independently by the increased
statistical significance of the ``dip'' in the ICL PNLF, and led to improved constraints on the
spatial distribution of the ICL thanks to better spatial coverage and statistics.  The ICL radial
surface density distribution is now consistent with a power law $I_{ICL} \propto R^{-\alpha}$, with
$\alpha_{\rm ICL}=0.79\pm0.15$, which is shallower than the Sersic profile for the smooth M87 halo. The
different PNLFs and spatial distributions confirm and strenghten the assessment by
\citet{longobardi15a} that the M87 halo and ICL are distinct components.  We note that the
transition from M87 to ICL is relatively sudden both in surface brightness and in kinematics
(LOSVD). Such sharp transitions are not expected in relaxed clusters. For example, around M49 in the
Virgo subcluster B, the BCG plus IGL system displays a continuous radial transition in both
kinematics and stellar population properties \citep{Hartke18}.

The dynamical properties of the ICL around M87 can be used as a benchmark for advanced
hydrodynamical simulations of galaxy clusters. The separation of central galaxy (M87) and ICL on the
basis of different LOSVDs has similar aspects as the classification as function of binding energy of
stellar particles in simulated cluster centers \citep{dolag10,cui14}. Stars with high binding
energies come from mergers of fairly massive progenitors, i.e.\ relaxation and merging processes that
led to rapid changes of the gravitational potential. As a result these particles have lost memory
of their progenitors, while particles with low binding energies still reflect the dynamics of their
lower mass satellite progenitors. In particle tagging methods \citep{cooper14} BCGs and ICL have
thus been associated with relaxed/unrelaxed accreted components.

The Illustris TNG simulations \citep{Pillep18} made detailed predictions on ICL fractions and
spatial distributions that can be compared with the results from the current investigation. In what
follows we assume a total halo mass of $\simeq 3\times 10^{14} M_\odot$ for the Virgo cluster
\citep{karachentsev2010}.  For this halo mass, the Illustris TNG simulations predict a best-fitting
power-law slope of $-\alpha_{\small\rm TNG,ICL} \sim -2.2$ to the 2D stellar mass surface density of
the combined halo and ICL. In the outer regions of M87 ($\rm{R<150\, kpc}$) we measure $-\alpha \simeq
-(2.0-2.5)$, in approximate agreement. The ICL alone has a shallower radial profile there,
$-\alpha_{\small\rm ICL} \simeq -0.8$.

For the Virgo cluster halo mass, the simulations predict approximate ICL stellar mass fractions out
to the virial radius of $\sim 0.35$ for an aperture $> 30$ kpc, and of $\sim 0.2$ for an aperture
$>100$ kpc \citep[][their Fig.~10]{Pillep18}. From \citet{longobardi15a}, in the radial range $\rm{7\,
kpc < R < 150\, kpc}$ the V-band luminosities of the M87 halo and ICL are $L_{\rm halo} =
4.42 \times 10^{10} L_\odot$ and $L_{\rm ICL} = 0.53 \times 10^{10} L_\odot$,
respectively\footnote{These luminosities account for the fraction of ICPNs in the range of
  velocities covered by the M87 halo}. For the M/L ratios, $\gamma^*$, we adopt values assigned by the
color-mass-to-light-ratio relation \citep{McGaugh14}, using the $B-V$ colours measured for the outer
halo and ICL, respectively. For the M87 outer halo, we adopt $\gamma^*_{V, \mathrm{M87}} =2.3$ for a
color of $B-V =0.76$, as in \citet{longobardi15b}; for the ICL, $\gamma^*_{V, \mathrm{ICL}} =1.0$
for a color $B-V \sim 0.6$ measured at $130$ kpc from the centre of M87 \citep{mihos17}. These
luminosities and M/L ratios result in a stellar mass fraction of $0.05$ for the ICL for $\rm{R<150\,
\mbox{kpc}}$ around M87. This appears lower than the predicted values but the comparison depends on
how quickly the slope of the ICL density steepens outside our observed range.

The Illustris TNG simulations also make predictions on the minimum progenitor stellar mass, such
that satellites of this mass and higher contribute 90\% of the total ex-situ stellar mass around
BCGs. For the $B-V \sim 0.6$ color of the M87+ICL light at 130 kpc distance \citep{mihos17} and a 10
Gyr old stellar population, the metallicity is [Fe/H]$\leq -1.0$. This agrees with the results of
\citet{williams07} who found from HST data in a Virgo ICL field that about 70-80\% of the stars have
ages $>10$~Gyr and mean metallicity $-1.0$. Using the stellar mass metallicity relation
\citep{zahid17}, we can infer the progenitor stellar mass for these stars, resulting in a few
$\times 10^{8} M_\odot$. By contrast, Fig.\, 13 from \citet{Pillep18} predicts larger progenitor
masses of a few $\times 10^{9} M_\odot$. The reason for this discrepancy is probably {\sl not}
related to the young dynamical age of the Virgo cluster core inferred from its unrelaxed velocity
distribution (see Sect.\ref{sec2} and Fig.~\ref{KS_test}).  While in this case the effective mass of
the accreting M87 subcluster might be a few times lower than the Virgo cluster virial mass, the
predicted minimum progenitor masses are insensitive to such variations. A more likely possibility
is that the simulations miss a population of lower-mass galaxies \citep{Hartke18}.

%\subsection{The kinematics of the outer regions of M87 and its relation with the X-ray potential}

\subsection{The smooth halo as tracer of  the gravitational potential in M87: comparison between optical and X-ray circular velocity curves}

After subtracting the strongly non-Gaussian ICL LOSVD, and the LOSVD of the crown substructure, the
remaining smooth M87 halo has kinematics centered on the galaxy's systemic velocity
($1307 \mathrm{kms}^{-1}$) and characterized by approximately Gaussian LOSVDs (see Section~\ref{sec2}
and Appendix~\ref{apx_histo}). Mean rotation velocities in the halo are $\lta 25 \mathrm{kms}^{-1}$
(Section~\ref{sec3}), and the velocity dispersion profile, after rising slowly to
$\sigma\simeq 300\, \mathrm{kms^{-1}}$ at $\mathrm{R_{avg} \simeq 50-70\, kpc}$, then declines steeply
down to $\sigma = 100\, \mathrm{kms^{-1}}$ at $\mathrm{R_{avg}\sim 135\, kpc}$
(Section~\ref{sec_sigma_R}; corresponding major-axis radii are 1.3 times larger).

By smooth halo we mean the part of the halo that is approximately phase-mixed (i.e., has
approximately Gaussian LOSVDs centered about the systemic velocity), at the resolution of our PN
survey. This is in contrast to the ICL component around M87, which is obviously non-Gaussian
(Fig.~\ref{KS_test}), and to the more localized phase-space substructure identified as the crown
\citep{longobardi15b}. However, because of the likely accretion origin of also the smooth outer
halo, and the long associated phase mixing timescales at $\rm{R\sim 100\, kpc}$, we expect that this
component too would show lower mass or amplitude substructures if it was possible to look at its
phase-space with a substantiallty larger number of stellar tracers. Nonetheless the working concept
of the smooth halo is useful because the approximately Gaussian LOSVDs enable us to determine
well-defined velocity dispersions and tracer densities for carrying out a Jeans analysis of the mass
and anisotropy at large radii.

M87 is an X-ray bright elliptical galaxy and the gravitational potential can be traced directly by
modeling the hot gas atmosphere, under the assumption of hydrostatic equilibrium
\citep[e.g.][]{nulsen95}. However, the comparison between the circular speed profiles computed from
dynamical modeling of the stars' LOS velocities and from the X-ray data in several nearby massive
ellipticals including M87 has indicated that the depth of the potential well derived from the X-ray
emitting hot gas is systematically lower than the corresponding optical value (from stars) such that
$v_{c,opt} = \eta \times v_{c,X}$ with $\eta = 1.10 - 1.14$ \citep{churazov10}. This implies that
the mass estimates from X-ray data underestimate the enclosed total mass by $21\%$ to $30\%$, and
has been considered as evidence for a significant non-thermal pressure support
\citep{Churazov08,gebhardt09,shen10,das10}.

Nonetheless we were able in Sect.~\ref{Xray} to obtain a consistent interpretation of the tracer
density and velocity dispersion profile of the smooth halo in the gravitational potential obtained
from the X-ray data of \citet{churazov10} and \citet{simionescu17}. Using the local, spherically
symmetric Jeans analysis method of \citet{churazov10} we found that the radial variation of the LOS
dispersion profile $\sigma(R_{\rm avg})$, rising from 270 to 300~kms$^{-1}$ in the radial range
$\mathrm{10\, kpc \le R_{\rm avg} \le 70\, kpc}$ followed by a decline to 100~kms$^{-1}$ at
$\rm{R_{avg}=135}$ kpc, can be reproduced by an isotropic stellar orbital distribution in the radial
range up to $\sim 60$ kpc, which becomes strongly radially anisotropic outside $70$ to $135$ kpc.
The strong decline at the largest radii is similar to what is measured in our own Milky Way halo
\citep[see Fig.15 in][]{BHG}.  The strong radial dependence of $\sigma(\rm{R_{avg}})$ for the smooth
M87 halo can be generated from a flat and then rising $v_{c,X}$, a steeper I(R) for the smooth M87
halo, as shown in Panels~A and B of Figure~\ref{circ_v}, and a varying orbital anisotropy profile
with radial dependence indicated in Panel~C of Figure~\ref{circ_v}.  This simplified picture
provides a consistent description of the velocity dispersion profile for the smooth M87 halo, and
sets the basis for a more sophisticated dynamical model to follow.

We also computed the predicted $\sigma_{\rm ICL}$ profile from the X-ray circular velocity profile
$v_{c,X}$ and the full Sersic profile (n=11 including the ICL; from K09) for an isotropic orbital
distribution.  Even with an increasing circular speed and a flatter surface brightness profile, the
predicted $\sigma(\rm{R_{avg}})$ at 135 kpc is $\sim 540$ kms$^{-1}$, i.e. it does not rise fast
enough to reproduce the upward $\sigma$ profile obtained when the ICL PNs are included (magenta full
dots in Fig.~\ref{sigma_profile}).  This is because of the non-Gaussian ICL LOSVD, and supports
previous assessments that the ICL around M87 is not (yet) in dynamical equilibrium.

We note that the modelling of the smooth halo indicates very strong radial anisotropy at the
outermost radii probed by the PN data. This suggests that if the hydrostatic interpretation of the
X-ray data significantly underestimated M87's circular velocity curve at $\sim 100 \mathrm{kpc}$
radius, it would be difficult to find a dynamical equilibrium model matching the low dispersion
there. Thus non-thermal pressure contributions may in fact be small at those radii, and the
hydrostatic pressure of the X-ray emitting gas therefore trace the enclosed mass.  We also note that
the new modelling obviates the need for a truncation of the density inferred by \citet{doherty09}.
This is because the new velocity dispersion profile and surface density profile of the smooth halo
appear consistent with the mass distribution from X-rays.  This is ultimately due to the better
statistics in the new ICPN data which allowed us to subtract the (non-equilibrium) ICL component
from both the surface brightness and LOS velocity dispersion data.

\subsection{The M87 kinematics as traced by GCs and ultra compact dwarfs}\label{newsec5.3}

In addition to PNs, GCs and UCDs are used as bright tracers to measure
LOSVs and thus overcome the limits represented by the very low surface
brightnesses characteristic of the outer regions of the M87 halo. It
is of interest then to compare the results from the different tracers in
order to assess any similarities or discrepancies, and understand the
origin of the latter.

\citet{strader11} presented a detailed kinematic analysis\footnote{We
  note that \cite{strader11} compute an extended comparison with
  previously published GC samples, and refer the reader to their work
  for further information.} of a sample of $\sim$400 GC' LOSVs that
covers the M87 halo out to 40\arcmin ($\sim 175\, \rm{kpc}$). They
found that all the GC populations are characterized by rotation that
becomes stronger at large radii. However GC subsamples with different
average colors have rotation that differs both in amplitude and
direction.

By comparing \citet{strader11} results to the PN kinematics of the M87
smooth halo presented in this study, we found differences that are
most significant with respect to the kinmatics of the metal-poor GC
subsample.  The fiducial velocity dispersion profile of the smooth M87
halo as traced by the stars and PNs is in broad agreement with the red
GC population, within the uncertainties (see also
Fig.~\ref{sigma_profile}). Differently from the red GCs subsample
though, the PN velocity field for the M87 smooth halo did not show any
signatures of rotation along the photometric minor axis, see
discussion in Sect.~\ref{sec3}.  For distances larger than 10\arcmin,
the bluer GC population and the M87 smooth halo PNs are rotating about
the galaxy's photometric minor axis, with the former having larger
amplitude of rotation.

The different kinematics shown by the different GC populations, may be
related to a Virgo ICGC population, in addition to a M87 GC halo
population.  \citet{durrell14} provided evidence for a Virgo ICGC
populations on the bases of an excess of number counts in the Virgo
core within in the extended area surveyed by Next Generation Virgo
Cluster Survey. Both \citet{durrell14} and \citet{ko17} found an ICGC
population mostly associated with blue GCs. This is consistent with
results from simulations \citep[e.g.][]{ramos15}, the latter showing that galaxies moving into
Virgo-like clusters are stripped mainly of their blue GC component
\citep[see also][for the first
  definitive kinematic detection of the ICGC population in the Virgo
  cluster]{longobardi18a}. We can then speculate that the more metal-poor
GCs contain a large fraction of IC population with its distinct
kinematics. It is interesting to notice that the LOSVD
associated to the 'green' GCs, as identified by \citet{strader11},
shows a secondary peak in their distribution (see Fig.~23 in their
work) similar to what is identified as the $v_{II}\sim 900\, \rm{kms^{-1}}$
component in the PN M87 halo sample, and later associated to the Virgo
ICL in Sect.~\ref{sec2}. If there is a fraction of red ICGCs, we expect
it to be lower.

A more recent study presented by \cite{zhang15} analysed the
properties of a sample of UCD galaxies within $2$ square degrees
centred on M87, and compared it to the properties of the red and blue
GC population.  Their results show that the surface number density
profile of the the UCDs is shallower than that of the blue GC sample
in the inner 15\arcmin and becomes as steep as the red GC component at
larger radii. Moreover they showed that the entire UCD system presents
a larger amplitude rotation than the GCs, with a rotational axis that
is more aligned with the red GC population in the same radial range.
These results show that the UCD system around M87 is kinematically
distinct from the GC population, and also from the M87 halo PNs. The
presence of distinct populations of tracers with multi-spins and
different kinematics can be understood in terms of an extended mass
assembly of the M87 halo. Tidal interactions and different specific
frequencies of the tracers, depending on the progenitor satellites
galaxies, can explain the occurrence of the kinematical diversity in
the halo of bright ellipticals like M87 \citep{coccato13}.

%CONCLUSIONS

 \section{Conclusions} 
\label{conclusions}

In this work we analysed the kinematics of 298 PNs in the outer regions of M87, covering the galaxy
halo and intracluster stars out to average radius $\rm{R_{avg}}\sim$ 135 kpc (corresponding to $\sim$
170 kpc along the major axis).  Our main results are:

(i) Including a newly identified $\rm{v_{II}\simeq 900\, kms^{-1}}$ component, the intracluster
stars have a strongly non-Gaussian LOSVD with a peak at that velocity, and strong, asymmetric
wings. The shape of the LOSVD is consistent with the LOSVD of the galaxies in the Virgo subcluster
A, and indicates that the ICL stars around M87 as well as the subcluster A galaxies are not (yet) in
dynamical equilibrium, signalling the on-going build-up of the Virgo cluster.

(ii) The so-called ``dip'' in the intracluster PN luminosity function has strengthened with respect to
earlier analysis, while no ``dip'' is seen in the PNLF of the M87 halo. This independently supports the
kinematic classification of the PNs into halo and ICL.  The surface density profile of the
kinematically tagged ICPNs decreases as a power law with radius, with negative logarithmic slope
$-\alpha_{\rm ICL}=-0.79\pm0.15$ in this region.

(iii) Based on the previously published B-V colour \citep{mihos17} and on resolved HST photometry
\citep{williams07}, the metallicity of the ICL population is estimated as [Fe/H]$\simeq -1.0$. This
suggests masses of a few $\times 10^{8} M_\odot$ for the ICL progenitor galaxies, which is an order
of magnitude less massive than the predictions from the Illustris TNG simulation \citep{Pillep18}.

(iv) The PNs in the smooth M87 halo, i.e.\ the part of the halo which is approximately phase-mixed
at the resolution of our PN survey, thus have a somewhat steeper surface density profile than the
total surface brightness profile from \citet{kormendy09} for halo and ICL together. The rotation of
these stars in the outer halo is small, $\lta 25\, \mathrm{kms}^{-1}$, safely in the slow rotator
regime.  The velocity dispersion profile of the smooth halo PNs rises slowly from the
$\sigma\simeq 270\, \mathrm{kms^{-1}}$ at $\simeq2-10\, \mathrm{kpc}$ seen in integrated spectra to
$\sigma\simeq 300\pm50 \, \mathrm{kms^{-1}}$ at average ellipse radii
$\mathrm{R_{avg} \simeq50- 70\, kpc}$, but then declines steeply down to
$\sigma = 100\, \mathrm{kms^{-1}}$ at $\mathrm{R_{avg}\sim 135\, kpc}$.

(v) Simple dynamical models indicate that the surface density and velocity dispersion profiles of
the smooth halo PN tracers at these large radii are consistent with being in approximate dynamical
equilibrium in the gravitational potential inferred from hydrostatic analysis of the X-ray emitting
gas. The X-ray circular velocity curve rises steeply outside $\sim 30$ kpc, reaching
$v_{c,X}\sim 700 \mathrm{kms}^{-1}$ at 200 kpc.  This requires the anisotropy of the halo stellar
orbits to change from an approximately isotropic distribution in the radial range up to $\sim 60$
kpc, to a strongly radially anisotropic configuration at the largest radii probed, $\rm{R_{avg}}=135$
kpc, as may be expected if the outer halo was accreted from infalling satellites.

\begin{appendix}

\section{LOSVD of the M87 outer regions}
\label{apx_histo}
\begin{figure*}[!h]
\centering \includegraphics[width=20.cm,clip=true, trim=4.cm 3cm 2.cm
  3cm]{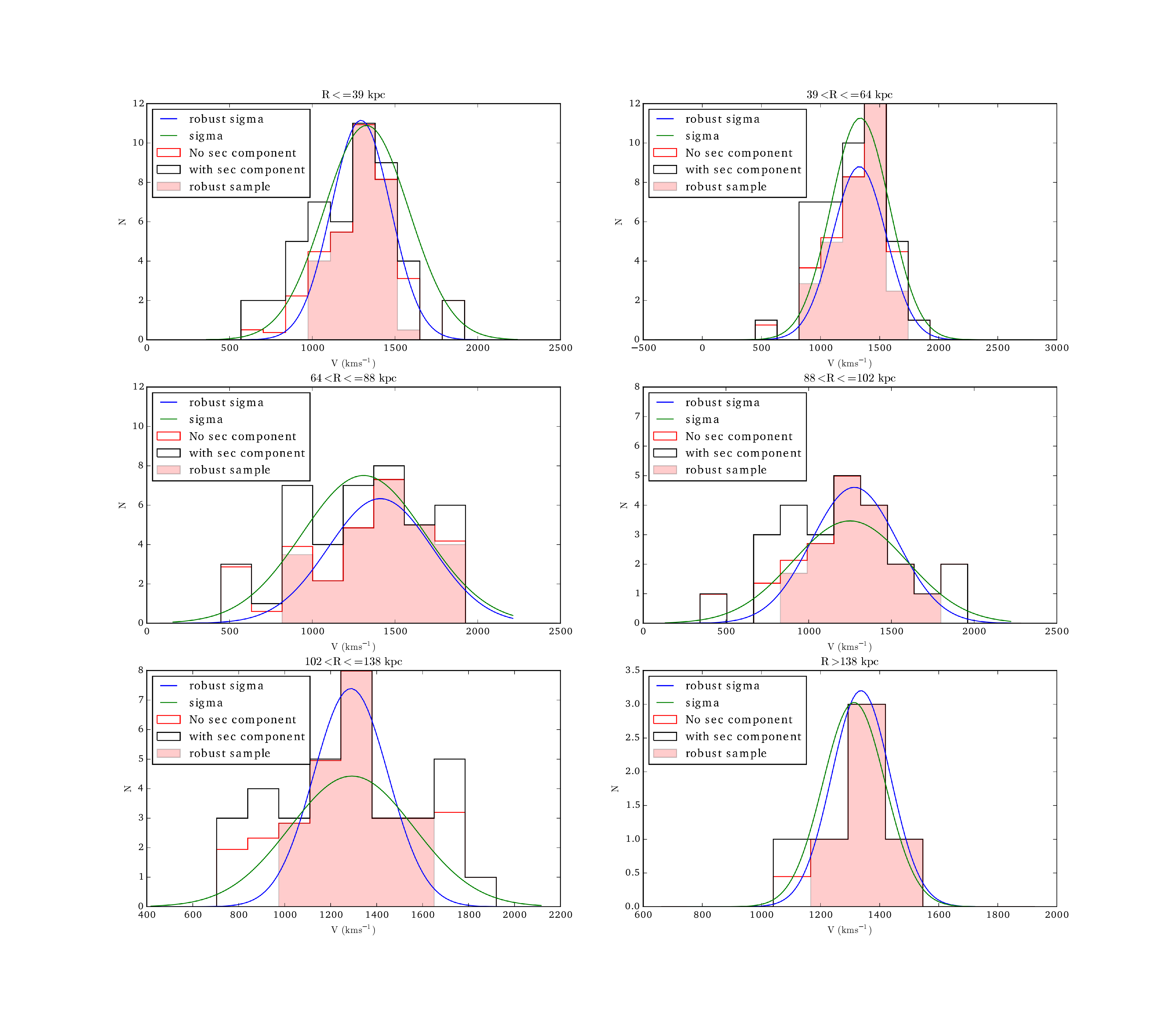}
   \caption{\small{ The PN LOSVDs in six radial bins from the M87's
       centre; the radial range is displayed on top of each panel. In
       each panel, the histogram limited by the black continuous line
       shows the LOSVD of the M87 halo PNs as classified in
       \cite{longobardi15a}. The histogram limited by the continuous
       red line shows the PN LOSVDs of the {\it smooth M87 halo}, once
       the $v_{II} \simeq 900$ kms$^{-1}$ component is accounted
       for. For smooth M87 halo PN LOSVDs we further plot two
       Gaussians, one with mean and dispersion computed as simple mean
       and standard deviation of the LOSV data (green continuous line)
       and that whose mean and dispersion value from from the robust
       estimator \citep[][blue continuous
       line]{mcneil10,longobardi15a}, and 2) . The red shaded histogram shows the PN LOSVDs
       once the robust estimator is applied. We note that in the
       fourth and fifth radial bin the velocity dispersion estimates
       deviate the most, because of the deviation of LOSVDs from a
       Gaussian and the limited statistic.}} \label{histograms}
\label{V_hist}
 \end{figure*}

This Appendix provides additional information the fiducial $\sigma$
profile and its dependencies on the uncertainties associated with {\it
  i)} the identification of the ICL stars in the range of velocities
associated with the M87 halo, {\it ii)} limited number statistics, and
{\it iii)} deviations of the LOSVD in radial bins from a Gaussian
distribution. In Figure~\ref{V_hist}, we show the PN LOSVDs in six
radial bins: they are those adopted in Fig.\ref{sigma_profile} and the
$\Delta R$ is indicated on the top of each panel. In each panel, we
illustrate the modification of the LOSVDs, depending on application of
the different constraints.

In each panel, we plot the histogram of the PN LOSVD associated with
the M87 halo by \cite{longobardi15a} with the black continuous line;
the histogram delimited by the red continuous line is computed for the
PN LOSVs without the $v_{II}\simeq 900$ kms$^{-1}$ component, and
represents the smooth M87 halo. The PN LOSVDs clearly deviate from a
straight Gaussian distribution and are affected by limited number
statistic in the outermost bins. Hence we proceed to estimate the the
second moment of the PN LOSVDs for the smooth M87 halo as a fiducial range,
whose limits are given by the values obtained from the standard deviation from
the measured LOSVs (upper limit) and from a robust sigma estimate 
from the LOSVD (lower limit). 

In each panel, we show two Gaussians profiles, one with mean and
dispersion values obtained from a robust procedure (blue continuous
line), and the second with mean and dispersion computed as simple mean
and standard deviation of the LOSVs in the bin (green continuous
line). The value obtained from the robust estimate is computed
according to \cite{mcneil10} and \cite{longobardi15a}.  The red
shaded histogram presents the PN LOSVDs selected by appling the robust
estimator in each bin.

The range of values for the second moment of the PN LOSVDs associated
with the smooth M87 halo in the different radial bins is shown in
Fig.\ref{sigma_profile} as gray shaded area function of the distance,
and represents our fiducial velocity dispersion profile.

  \section{PN catalogue}
Here we present the M87 PN catalogue obtained from the PN photometric
and spectroscopic surveys carried out with Suprime-Cam@Subaru and
FLAMES@VLT, respectively, and presented in \citet{longobardi13} and
\citet{longobardi15a}. In this catalogue we provide the PN coordinates
(J2000 system), the [OIII]$\lambda5007$\AA\ magnitudes, and
heliocentric $\rm{V_{LOS}}$ measured from a Gaussian fit to the
[OIII]$\lambda$5007\AA\ emission. In high S/N spectra, we also detect the
redshifted [OIII] $\lambda$4959/5007\AA\ doublet. Typical S/N ratios
for the spectroscopically confirmed PN [OIII]]$\lambda$5007\AA\ cover
  a range of $2.5 \lesssim \mathrm{S/N}\lesssim 15.0$ per resolution
  element. From the repeated observations of the same candidates in
  different FLAMES plate configurations we estimated the median
  deviation of velocity measurements to be 4.2 $\mathrm{kms^{-1}}$,
  and the hole distribution covers a range of $0.6 < \Delta \mathrm{V}
  < 16.2 \mathrm{kms^{-1}}$. {\rm In case of repeated observations the 
given heliocentric velocity has been estimated from the spectrum with the highest S/N.} 
 Longobardi et al 2015a discussed a
    statistical approach to determine the fraction of misclassified
    PNs based on the analysis of stacked PN spectra. They determined
    that 2\% of the entire sample (7 PNs) could represent
    misclassified spectra \citep[for more details
      see][]{longobardi15a}.

The table is divided in three parts: 1) PNs that have higher
probability of belonging to the smooth halo component and PNs that
have high probability to belong to the 'crown' structure (indicated by
an *) \citep{longobardi15b}, 2) PNs that have higher probability to
belong to the additional ICL component as determined in this work (see
Sect.~\ref{sec2}, and 3) PNS that have been assigned to the ICL
component by \citet{longobardi15a}).

\begin{longtable}[!h]{ccccccccc}
  \caption{\small{Spectroscopically confirmed PNs. Column 1: PN identifier according to the IAU regulations.  Column 2: Field ID following conventions in \citet{longobardi15a}. FCJ and F7 are data from \citet{doherty09}. Column 3 \& 4: Right Ascension and Declination. Columns 5: Line-of-sight velocity corrected for heliocentric velocity.  Column 6: Measured $m_{5007}$ magnitudes from \citet{longobardi13}. Column 7: probability to belong to the main M87 halo. Column 8: Flag indicating the detection (yes) or not (--) of the [OIII] $\lambda$4959/5007\AA\ doublet. Column 9: S/N per resolution element for the redshifted [OIII] $\lambda$5007\AA\ emission line.}  \label{Halo_table}} \\
%\label{halo_table}

\hline\hline
\textbf{PN ID} &\textbf{Field} & \textbf{RA} & \textbf{DEC} & \textbf{$V_{\mathrm{LOS}}$} & \textbf{$\mathrm{mag_{5007}}$} & \textbf{$\Gamma_{i}$ }& \textbf{ [OIII] Doublet} & \textbf{S/N}\\
& & J2000 & J2000  &  &  &\\
& & (deg) & (deg)  & $\rm (kms^{-1})$ &  &\\ \hline

\endfirsthead
\multicolumn{9}{c}%
{\bfseries \tablename\ \thetable\ -- \textit{continued from previous page}} \\
\hline\hline
\textbf{PN ID} &\textbf{Field} & \textbf{RA} & \textbf{DEC} & \textbf{$V_{\mathrm{LOS}}$} & \textbf{$\mathrm{mag_{5007}}$} & \textbf{$\Gamma_{i}$ }& \textbf{ [OIII] Doublet} & \textbf{S/N} \\
& & J2000 & J2000  &  &  & & &\\
& & (deg) & (deg)  & $\rm (kms^{-1})$ &  & & &\\ \hline\endhead

\hline
\multicolumn{9}{r}{\bfseries \tablename\ \thetable\ -- \textit{continued on next page}} \\
\hline
\endfoot

\hline
\hline
\endlastfoot

M87PN J123033.96+123050.0 & FCJ & 187.6415 & 12.5139 & 1467.9 & 26.9 & 1.0 & \rm{yes} & 4.2 \\
M87PN J123041.30+123226.1 &FCJ & 187.6721 & 12.5406 & 940.7 & 26.2 & 1.0 &\rm{yes} & 11.2\\
M87PN J123024.36+123302.8 & FCJ & 187.6015 & 12.5508 & 1109.0 & 26.8 & 1.0 &\rm{yes} & 5.5\\
M87PN J123034.80+123605.0 &  FCJ & 187.6450 & 12.6014 & 1390.1 & 27.3 & 1.0 &-- & 4.3\\
M87PN J123113.36+123318.7 &FCJ & 187.8057 & 12.5552 & 1277.4 & 27.0 & 1.0 &\rm{yes} & 4.1\\
M87PN J123053.78+123826.8 &FCJ & 187.7241 & 12.6408 & 1743.6 & 26.9 & 1.0 &\rm{yes} & 8.2\\
M87PN J122932.04+124453.1 & F71 & 187.3835 & 12.7481 & 1223.5 & 29.8 & 1.0 &\rm{yes} & 10.9 \\
M87PN J122917.01+125203.7 &F71 & 187.3209 & 12.8677 & 1230.1 & 29.2 & 1.0 &\rm{yes} & 11.3\\
M87PN J122907.32+125043.0 & F71 & 187.2805 & 12.8453 & 1302.6 & 29.2 & 1.0 &\rm{yes} & 13.0 \\
M87PN J122856.23+123632.0 &F71 & 187.2343 & 12.6089 & 1415.8 & 29.0 & 1.0 &\rm{yes} & 13.5\\
M87PN J122847.28+123527.2 & F71 & 187.1970 & 12.5909 & 1314.0 & 29.4 & 1.0 &\rm{yes} & 12.7\\
M87PN J123052.44+122113.3 & FC & 187.7185 & 12.3537 & 1009.6 & 26.2 & 0.8 &\rm{yes} &7.9 \\
M87PN J123053.13+122055.6 &FC & 187.7214 & 12.3488 & 1413.4 & 27.1 & 1.0 &\rm{yes} & 6.6\\
M87PN J123042.36+122538.2 &FC&187.6765& 12.4273 &795.7 &27.2& 1.0 &\rm{yes} & 6.1\\
M87PN J123052.48+122539.0 &   FC & 187.7187 & 12.4275 & 1256.5 & 27.0 & 1.0 &\rm{yes} &8.2 \\
M87PN J123050.47+122006.0 &   FC & 187.7103 & 12.3350 & 1243.8 & 27.1 & 1.0 &\rm{yes} & 6.6\\
M87PN J123039.52+122335.5 &FC & 187.6647 & 12.3932 & 930.5 & 27.4 & 0.6 &\rm{yes} & 6.3\\
M87PN J123100.00+122122.6 &    FC & 187.7500 & 12.3563 & 1362.5 & 27.4 & 1.0 &\rm{yes} &8.7 \\
M87PN J123051.48+121953.4 &  FC & 187.7145 & 12.3315 & 1355.9 & 26.9 & 1.0 &\rm{yes} & 6.9\\
M87PN J123051.55+121940.4 &FC & 187.7148 & 12.3279 & 1425.4 & 26.7 & 1.0 &\rm{yes} & 5.7\\
M87PN J123051.43+122654.6 & FC & 187.7143 & 12.4485 & 1285.2 & 27.7 & 1.0 &\rm{yes} & 5.0\\
M87PN J123036.91+122407.5 &FC & 187.6538 & 12.4021 & 1204.3 & 27.4 & 0.9 &\rm{yes} & 5.5\\
M87PN J123036.84+122617.8 & F1-F03  & 187.6535 & 12.4383 & 1264.8 & 26.8 & 1.0 &\rm{yes} &9.0 \\
M87PN J123059.28+121927.4 & FC & 187.7470 & 12.3243 & 1233.1 & 27.6 & 1.0 &\rm{yes} & 5.2\\
M87PN J123046.08+122753.2 & FC & 187.6920 & 12.4648 & 1258.2 & 26.6 & 1.0 &\rm{yes} & 8.6\\
M87PN J123046.08+121937.5 & FC & 187.6920 & 12.3271 & 1398.4 & 26.6 & 1.0 &\rm{yes} & 8.0\\
M87PN J123035.20+122617.1 &FC & 187.6467 & 12.4381 & 1032.3 & 26.9 & 0.6 & \rm{yes}& 6.0\\
M87PN J123034.99+122414.4 & FC & 187.6458 & 12.4040 & 1338.5 & 26.5 & 1.0 &\rm{yes} & 8.7\\
M87PN J123047.35+121840.3 &FC & 187.6973 & 12.3112 & 1057.5 & 26.3 & 0.9 &\rm{yes} & 6.5\\
M87PN J123042.12+122914.2 & F1-F03  & 187.6755 & 12.4873 & 1643.5 & 26.7 & 1.0 &\rm{yes} &5.6 \\
%M87PN J123042.12+122914.2 & FC & 187.6755 & 12.4873 & 1641.1 & 26.7 & 1.0 &\rm{yes} & 5.6\\
M87PN J123032.68+122700.3 &FC & 187.6362 & 12.4501 & 1444.6 & 27.8 & 1.0 &-- & 5.2\\
M87PN J123052.27+121744.8 &FC & 187.7178 & 12.2958 & 1305.0 & 27.0 & 1.0 &\rm{yes} & 4.8\\
M87PN J123033.96+122850.8 & FC & 187.6415 & 12.4808 & 1473.3 & 27.2 & 1.0 &\rm{yes} & 5.5\\
M87PN J123032.66+122222.4 &FC & 187.6361 & 12.3729 & 1647.7 & 27.4 & 0.6 & \rm{yes}& 5.5\\
M87PN J123037.53+122954.6 & F1-F03  & 187.6564 & 12.4985 & 1204.3 & 27.3 & 1.0 &-- &6.7 \\
M87PN J123108.23+121859.4 & F2-F02  & 187.7843 & 12.3165 & 967.1 & 27.7 & 0.4 &\rm{yes} & 4.9\\
M87PN J123029.28+122613.2 &  F1-F03  & 187.6220 & 12.4370 & 1335.5 & 27.9 & 1.0 & \rm{yes}& 8.2\\
M87PN J123042.52+121829.5 &FC & 187.6772 & 12.3082 & 1325.3 & 27.5 & 1.0 &\rm{yes} & 6.2\\
M87PN J123035.61+123022.3 &F1-F03  & 187.6484 & 12.5062 & 1472.1 & 27.0 & 0.6 &\rm{yes} &8.3 \\
M87PN J123107.63+121731.5 &F2-F02  & 187.7818 & 12.2921 & 1618.3 & 26.6 & 0.5 &-- & 3.8\\
M87PN J123100.98+122742.1 &FC & 187.7541 & 12.4617 & 1846.6 & 27.8 & 1.0 &\rm{yes} & 6.0\\
M87PN J123111.25+122117.2 &F2-F02  & 187.7969 & 12.3548 & 1389.5 & 27.0 & 0.9 & \rm{yes}& 5.2\\
M87PN J123105.37+121639.7 &F2-F02  & 187.7724 & 12.2777 & 1119.8 & 28.3 & 1.0 & \rm{yes}& 5.3\\
M87PN J123037.12+121908.0 &FC & 187.6547 & 12.3189 & 1938.3 & 26.7 & 1.0 &\rm{yes} & 6.5\\
M87PN J123028.41+122854.8 &F1-F03  & 187.6184 & 12.4819 & 1086.3 & 27.2 & 0.7 & \rm{yes}& 7.9\\
M87PN J123110.12+121748.1 &F2-F02*  & 187.7922 & 12.2967 & 1566.8 & 28.4 & 0.4 & \rm{yes}& 6.9\\
M87PN J123112.67+122015.3 &FC & 187.8028 & 12.3376 & 1883.8 & 26.6 & 1.0 &-- & 2.7\\
M87PN J123031.77+123039.6 & F1-F03  & 187.6324 & 12.5110 & 1485.3 & 27.1 & 0.6 &\rm{yes} & 7.6\\
M87PN J123113.2+121954.84 &  FC & 187.8050 & 12.3319 & 1354.7 & 26.7 & 1.0 & \rm{yes}& 6.6\\
M87PN J123026.16+122400.3 & F1-F03  & 187.6090 & 12.4001 & 1154.6 & 27.3 & 0.7 & \rm{yes}& 7.2\\
M87PN J123047.35+121557.2 &FC* & 187.6973 & 12.2659 & 944.3 & 28.0 & 0.3 & \rm{yes}& 5.6\\
M87PN J123024.91+122421.9 &FC* & 187.6038 & 12.4061 & 1584.8 & 27.7 & 0.4 & \rm{yes}& 5.7\\
M87PN J123027.38+122217.0 &F1-F03  & 187.6141 & 12.3714 & 1132.4 & 26.8 & 0.6 & \rm{yes}& 7.7\\
M87PN J123111.61+121625.3 &F2-F02  & 187.7984 & 12.2737 & 1482.9 & 26.7 & 0.6 & \rm{yes}& 7.3\\
M87PN J123048.19+123145.4 &F1-F01  & 187.7008 & 12.5293 & 1317.0 & 26.6 & 1.0 & \rm{yes}& 8.3\\
M87PN J123041.30+123226.1 &F1-F01  & 187.6721 & 12.5406 & 1309.2 & 27.1 & 1.0 & \rm{yes}& 7.9\\
M87PN J123027.33+123116.6 &F1-F03  & 187.6139 & 12.5213 & 1306.8 & 26.9 & 1.0 & \rm{yes}& 8.6\\
M87PN J123057.91+123031.6 &F1-F01*  & 187.7413 & 12.5088 & 1542.3 & 27.2 & 0.4 & \rm{yes}& 7.8\\
M87PN J123027.00+123128.9 &    F1-F03  & 187.6125 & 12.5247 & 1345.7 & 27.0 & 1.0 & \rm{yes}& 4.3\\
M87PN J123109.24+122708.6 & F1-F01  & 187.7885 & 12.4524 & 1243.8 & 27.5 & 0.9 & -- & 5.6\\
M87PN J123047.06+121455.3 &FC & 187.6961 & 12.2487 & 1121.6 & 27.3 & 1.0 & \rm{yes}& 5.3\\
M87PN J123117.40+122207.3 &  F2-F02  & 187.8225 & 12.3687 & 1419.4 & 26.9 & 0.9 & \rm{yes}& 6.3\\
M87PN J123020.30+122611.7 &F1-F03  & 187.5846 & 12.4366 & 1337.9 & 26.9 & 1.0 & \rm{yes}& 8.7\\
M87PN J123042.36+121532.7 & FC & 187.6765 & 12.2591 & 980.2 & 26.9 & 0.4 & \rm{yes}& 5.8\\
M87PN J123057.86+121324.6 & F2-F02  & 187.7411 & 12.2235 & 1478.7 & 27.9 & 0.6 & \rm{yes}& 6.0\\
M87PN J123019.82+122506.6 & F1-F02  & 187.5826 & 12.4185 & 1406.8 & 26.6 & 0.9 & \rm{yes}& 7.9\\
M87PN J123052.29+123231.2 & F1-F02  & 187.7179 & 12.5420 & 1499.1 & 27.0 & 0.5 & \rm{yes}& 7.4\\
M87PN J123025.44+123232.2 & F1-F03  & 187.6060 & 12.5423 & 1067.7 & 27.6 & 0.7 & \rm{yes}& 7.1\\
M87PN J123102.95+121301.9 &F2-F03*  & 187.7623 & 12.2172 & 1563.2 & 26.8 & 0.4 & \rm{yes}& 7.1\\
M87PN J123120.64+121800.0 &    FC* & 187.8360 & 12.3000 & 1604.5 & 27.0 & 0.5 & \rm{yes}& 7.8\\
M87PN J123020.61+123126.0 &F1-F03  & 187.5859 & 12.5239 & 1161.2 & 27.2 & 0.9 & \rm{yes}& 8.0\\
M87PN J123055.32+121250.4 &  F2-F03  & 187.7305 & 12.2140 & 1386.5 & 26.9 & 0.6 & \rm{yes}& 6.7\\
M87PN J123022.68+123243.4 & F1-F03  & 187.5945 & 12.5454 & 1276.8 & 27.4 & 0.5 & \rm{yes}& 6.6\\
M87PN J123108.11+121227.3 &F2-F02*  & 187.7838 & 12.2076 & 1070.1 & 27.6 & 0.4 & \rm{yes}& 6.9\\
M87PN J123037.94+123441.1 &F1-F02  & 187.6581 & 12.5781 & 1627.9 & 27.8 & 1.0 & \rm{yes}& 8.7\\
M87PN J123017.16+123038.1 & F1-F03  & 187.5715 & 12.5106 & 1142.6 & 27.1 & 0.8 & \rm{yes}& 7.0\\
M87PN J123123.83+122019.3 &FC* & 187.8493 & 12.3387 & 1316.4 & 28.3 & 0.4 & \rm{yes}& 4.4\\
M87PN J123101.34+123207.8 & FC & 187.7556 & 12.5355 & 988.6 & 26.9 & 0.5 & \rm{yes}& 7.0\\
M87PN J123119.27+122522.4 &FC* & 187.8303 & 12.4229 & 1337.9 & 27.8 & 0.4 & \rm{yes}& 5.9\\
M87PN J123038.52+123515.3 & F1-F02  & 187.6605 & 12.5876 & 1799.3 & 27.6 & 1.0 & \rm{yes}& 7.2\\
M87PN J123125.20+121823.7 &  F2-F02  & 187.8550 & 12.3066 & 1473.3 & 27.8 & 1.0 & \rm{yes}& 8.1\\
M87PN J123020.52+123341.7 & F1-F03*  & 187.5855 & 12.5616 & 1296.0 & 27.1 & 0.4 & \rm{yes}& 7.8\\
M87PN J123104.44+121120.7 & F2-F02* & 187.7685 & 12.1891 & 1097.7 & 27.5 & 0.5 & \rm{yes}& 6.7\\
M87PN J123021.36+123414.8 & F1-F03  & 187.5890 & 12.5708 & 1487.7 & 27.9 & 1.0 & \rm{yes}& 7.1\\
M87PN J123012.76+122758.6 &F1-F03  & 187.5532 & 12.4663 & 968.8 & 27.9 & 0.6 & \rm{yes}& 5.5\\
M87PN J123115.79+121213.3 &F2-F02  & 187.8158 & 12.2037 & 1402.7 & 27.6 & 0.8 & \rm{yes}& 7.4\\
M87PN J123033.26+121516.9 &FC* & 187.6386 & 12.2547 & 1349.9 & 28.2 & 0.4 & \rm{yes}& 5.4\\
M87PN J123051.24+123444.4 &  F1-F02  & 187.7135 & 12.5790 & 604.0 & 27.3 & 0.8 & \rm{yes}& 6.6\\
M87PN J123057.16+123346.0 &F1-F01  & 187.7382 & 12.5628 & 1274.4 & 27.8 & 0.5 & \rm{yes}& 8.2\\
M87PN J123040.32+121323.8 & FC & 187.6680 & 12.2233 & 1378.1 & 27.7 & 0.6 & --& 5.8\\
M87PN J123100.19+121059.1 &F2-F03  & 187.7508 & 12.1831 & 1514.7 & 27.8 & 1.0 & \rm{yes}& 5.5\\
M87PN J123027.72+123548.4 & F1-F03*  & 187.6155 & 12.5968 & 1357.7 & 26.6 & 0.5 & \rm{yes}& 4.7\\
M87PN J123042.88+123549.2 & F1-F01*  & 187.6787 & 12.5970 & 1297.2 & 27.8 & 0.4 & \rm{yes}& 7.3\\
M87PN J123031.60+123605.0 &F1-F03  & 187.6317 & 12.6014 & 1482.3 & 27.1 & 1.0 & \rm{yes}& 7.9\\
M87PN J123012.86+123150.1 &F1-F03  & 187.5536 & 12.5306 & 1202.5 & 27.5 & 1.0 & \rm{yes}& 7.2\\
M87PN J123102.92+121014.1 &F2-F03  & 187.7622 & 12.1706 & 1500.9 & 27.8 & 1.0 & \rm{yes}& 6.7\\
M87PN J123039.96+123637.8 &  F1-F01*  & 187.6665 & 12.6105 & 1333.7 & 27.3 & 0.4 & \rm{yes}& 6.2\\
M87PN J123058.24+121006.6 & F2-F03  & 187.7427 & 12.1685 & 1734.6 & 26.9 & 1.0 & \rm{yes}& 3.9\\
M87PN J123105.90+123249.9 &F1-F01  & 187.7746 & 12.5472 & 1712.4 & 27.0 & 1.0 & \rm{yes}& 9.2\\
M87PN J123013.17+122218.4 &F1-F03*  & 187.5549 & 12.3718 & 1288.2 & 27.5 & 0.4 & \rm{yes}& 6.8\\
M87PN J123127.91+121407.4 &F2-F02  & 187.8663 & 12.2354 & 1530.8 & 28.1 & 1.0 & \rm{yes}& 5.1\\
M87PN J123103.74+120945 &   F2-F03  & 187.7656 & 12.1625 & 1620.7 & 26.5 & 1.0 & \rm{yes}& 4.3\\
M87PN J123125.87+121245.7 &F2-F02  & 187.8578 & 12.2127 & 1383.5 & 27.0 & 0.6 & \rm{yes}& 8.4\\
M87PN J123053.59+123539.8 &F1-F01*  & 187.7233 & 12.5944 & 1049.7 & 28.2 & 0.4 & \rm{yes}& 6.3\\
M87PN J123103.52+123342.4 &F1-F01*  & 187.7647 & 12.5618 & 1360.7 & 27.6 & 0.5 & \rm{yes}& 9.6\\
M87PN J123049.60+121037.2 & F2-F03  & 187.7067 & 12.1770 & 1456.0 & 27.7 & 1.0 & \rm{yes}& 7.4\\
M87PN J123018.52+123609.3 &F1-F03  & 187.5772 & 12.6026 & 1648.9 & 26.8 & 1.0 & \rm{yes}&7.6 \\
M87PN J123052.94+121000.1 &F2-F03  & 187.7206 & 12.1667 & 618.3 & 27.9 & 0.9 & \rm{yes}& 6.7\\
M87PN J123101.15+123433.9 &F1-F01  & 187.7548 & 12.5761 & 955.7 & 26.7 & 0.6 & \rm{yes}& 9.6\\
M87PN J123007.75+123126.4 & F1-F03  & 187.5323 & 12.5240 & 1202.5 & 27.9 & 1.0 & \rm{yes}&5.7 \\
M87PN J123006.72+122833.6 &  F1-F03  & 187.5280 & 12.4760 & 1268.4 & 27.5 & 0.6 & \rm{yes}& 8.9\\
M87PN J123030.38+123747.2 &F1-F03*  & 187.6266 & 12.6298 & 1062.9 & 28.0 & 0.4 & \rm{yes}& 6.2\\
M87PN J123132.61+12164.08 &   F2-F02  & 187.8859 & 12.2800 & 1103.1 & 27.3 & 0.5 & \rm{yes}& 9.5\\
M87PN J123131.65+122104.6 &F2-F02  & 187.8819 & 12.3513 & 977.8 & 27.2 & 0.5 & \rm{yes}& 7.5\\
M87PN J123024.04+123735.7 &F1-F01  & 187.6002 & 12.6266 & 1264.8 & 27.5 & 0.6 & \rm{yes}& 9.5\\
M87PN J123015.24+123609.7 & F1-F03  & 187.5635 & 12.6027 & 1399.6 & 27.3 & 0.7 & \rm{yes}& 7.9\\
M87PN J123006.26+122606.0 &   F1-F02  & 187.5261 & 12.4350 & 1126.4 & 26.8 & 0.4 & \rm{yes}& 7.4\\
M87PN J123016.03+123656.5 &F1-F02*  & 187.5668 & 12.6157 & 1347.5 & 27.0 & 0.5 & \rm{yes}& 6.6\\
M87PN J123010.39+123538.4 & F1-F03*  & 187.5433 & 12.5940 & 1087.5 & 27.5 & 0.4 & \rm{yes}& 7.0\\
M87PN J123007.58+123421.0 &   F1-F03*  & 187.5316 & 12.5725 & 1261.8 & 28.0 & 0.2 & \rm{yes}& 6.6\\
M87PN J123116.56+120838.7 & F2-F02*  & 187.8190 & 12.1441 & 1098.3 & 27.5 & 0.4 & \rm{yes}& 6.6\\
M87PN J123007.72+122301.3 &F1-F03  & 187.5322 & 12.3837 & 1642.3 & 26.9 & 1.0 & \rm{yes}& 4.6\\
M87PN J123059.30+123558.2 & F1-F01  & 187.7471 & 12.5995 & 1421.2 & 26.8 & 1.0 & \rm{yes}& 6.5\\
M87PN J123042.88+121018.4 &F2-F03  & 187.6787 & 12.1718 & 695.0 & 27.4 & 0.6 & \rm{yes}& 5.1\\
M87PN J123109.57+123342.8 &F1-F01*  & 187.7899 & 12.5619 & 1245.0 & 27.3 & 0.2 & \rm{yes}& 8.1\\
M87PN J123135.35+121917.0 &F2-F02  & 187.8973 & 12.3214 & 986.2 & 27.1 & 0.7 & \rm{yes}& 8.8\\
M87PN J123015.07+123739.3 &F1-F01*  & 187.5628 & 12.6276 & 1322.3 & 27.9 & 0.3 & \rm{yes}& 7.7\\
M87PN J123050.83+123745.8 &F1-F01*  & 187.7118 & 12.6294 & 1228.9 & 28.9 & 0.3 & \rm{yes}& 7.6\\
M87PN J123019.46+123832.2 &F1-F01  & 187.5811 & 12.6423 & 1033.5 & 26.8 & 0.7 & \rm{yes}& 9.5\\
M87PN J123136.26+121502.8 &F2-F02  & 187.9011 & 12.2508 & 1427.8 & 27.0 & 1.0 & --& 4.5\\
M87PN J123031.60+123929.1 &F1-F02*  & 187.6317 & 12.6581 & 1331.3 & 27.6 & 0.4 & \rm{yes}& 7.3\\
M87PN J123123.16+122941.6 & F1-F01*  & 187.8465 & 12.4949 & 1117.4 & 27.2 & 0.2 & \rm{yes}& 8.8\\
M87PN J123113.36+123319.0 &F1-F01*  & 187.8057 & 12.5553 & 1285.8 & 27.3 & 0.2 & \rm{yes}& 7.0\\
M87PN J123043.20+123902.1 &  F1-F02  & 187.6800 & 12.6506 & 958.7 & 27.2 & 0.6 & \rm{yes}& 8.3\\
M87PN J123111.04+120711.6 & F2-F02  & 187.7960 & 12.1199 & 1509.3 & 26.6 & 1.0 & \rm{yes}& 3.5\\
M87PN J123000.33+123016.5 &F1-F02*  & 187.5014 & 12.5046 & 1237.3 & 27.5 & 0.3 & \rm{yes}& 9.1\\
M87PN J122959.88+123114.8 & F1-F03  & 187.4995 & 12.5208 & 1354.7 & 27.9 & 0.6 & \rm{yes}& 7.0\\
M87PN J123138.59+121435.8 &F2-F02  & 187.9108 & 12.2433 & 931.7 & 27.5 & 0.5 & \rm{yes}& 5.1\\
M87PN J123105.37+123606.1 &F1-F01  & 187.7724 & 12.6017 & 1467.4 & 27.6 & 1.0 & \rm{yes}& 7.0\\
M87PN J123011.49+123831.2 & F1-F01*  & 187.5479 & 12.6420 & 1083.3 & 27.6 & 0.4 & \rm{yes}& 7.0\\
M87PN J123043.32+120849.5 & F2-F03*  & 187.6805 & 12.1471 & 1287.0 & 26.6 & 0.2 & \rm{yes}& 6.3\\
M87PN J123059.56+123724.6 & F1-F01*  & 187.7482 & 12.6235 & 1224.1 & 26.8 & 0.3 & \rm{yes}& 5.2\\
M87PN J123053.78+123826.8 &F1-F01  & 187.7241 & 12.6408 & 1756.7 & 27.0 & 1.0 & \rm{yes}& 8.5\\
M87PN J123115.64+123330.6 & F1-F01*  & 187.8152 & 12.5585 & 1136.0 & 27.0 & 0.2 & --& 4.5\\
M87PN J123138.56+121304.0 &F2-F02  & 187.9107 & 12.2178 & 1618.3 & 27.7 & 1.0 & \rm{yes}&7.1 \\
M87PN J123049.15+123921.6 & F1-F02*  & 187.7048 & 12.6560 & 1146.8 & 27.0 & 0.4 & \rm{yes} & 7.4\\
M87PN J123028.94+121132.2 &FC & 187.6206 & 12.1923 & 506.3 & 26.8 & 1.0 & \rm{yes}& 5.9\\
M87PN J123125.12+120717.7 &F2-F02  & 187.8547 & 12.1216 & 925.1 & 27.8 & 0.5 & \rm{yes}& 7.4\\
M87PN J123140.75+121937.2 & F2-F02*  & 187.9198 & 12.3270 & 1259.4 & 27.9 & 0.2 & \rm{yes}& 6.3\\
M87PN J123110.94+120600.0 &   F2-F03  & 187.7956 & 12.1000 & 1524.9 & 27.2 & 1.0 & \rm{yes}& 6.3\\
M87PN J123140.12+122108.6 &F2-F02  & 187.9172 & 12.3524 & 457.8 & 27.3 & 1.0 & \rm{yes}& 7.4\\
M87PN J122957.69+123302.5 &F1-F03*  & 187.4904 & 12.5507 & 1317.6 & 27.1 & 0.3 & \rm{yes}& 8.5\\
M87PN J123024.12+124058.0 & F1-F02  & 187.6005 & 12.6828 & 1565.0 & 27.1 & 1.0 & \rm{yes}& 8.0\\
M87PN J123021.21+124103.8 &F1-F03*  & 187.5884 & 12.6844 & 1155.8 & 27.6 & 0.4 & --& 6.1\\
M87PN J123142.19+121354.4 &F2-F02*  & 187.9258 & 12.2318 & 1299.0 & 27.6 & 0.2 & \rm{yes}& 7.5\\
M87PN J123017.64+124050.8 & FEDGE  & 187.5735 & 12.6808 & 1863.4 & 28.2 & 1.0 & \rm{yes}& 5.0\\
M87PN J123049.15+124012.3 &F1-F01  & 187.7048 & 12.6701 & 1370.3 & 28.3 & 0.7 & \rm{yes}& 5.9\\
M87PN J122958.08+123534.8 &  F1-F02  & 187.4920 & 12.5930 & 1792.1 & 27.9 & 1.0 & \rm{yes}& 6.7\\
M87PN J123142.21+121219.8 & F2-F02*  & 187.9259 & 12.2055 & 1258.2 & 27.8 & 0.2 & \rm{yes}& 6.0\\
M87PN J122955.96+122607.8 & F1-F03*  & 187.4832 & 12.4355 & 1231.3 & 27.2 & 0.3 & \rm{yes}& 8.2\\
M87PN J122957.55+123626.2 &F1-F03  & 187.4898 & 12.6073 & 1599.2 & 27.6 & 1.0 & \rm{yes}& 9.3\\
M87PN J123102.25+123848.4 &F1-F01  & 187.7594 & 12.6468 & 1169.6 & 27.7 & 0.5 & \rm{yes}& 6.8\\
M87PN J122953.06+123233.3 &F1-F03*  & 187.4711 & 12.5426 & 1297.2 & 28.0 & 0.2 & \rm{yes}& 7.3\\
M87PN J122956.06+123636.0 &   F1-F03*  & 187.4836 & 12.6100 & 1223.5 & 27.1 & 0.3 & \rm{yes}& 7.2\\
M87PN J123025.29+124245.3 &F1-F02  & 187.6054 & 12.7126 & 1098.3 & 26.3 & 0.9 & \rm{yes}& 8.2\\
M87PN J122954.19+122448.2 &F1-F03  & 187.4758 & 12.4134 & 935.3 & 27.7 & 0.6 & \rm{yes}& 7.2\\
M87PN J123052.08+124106.7 & F1-F01  & 187.7170 & 12.6852 & 1920.3 & 27.3 & 1.0 & \rm{yes}& 6.7\\
M87PN J122954.91+123733.6 & F1-F03  & 187.4788 & 12.6260 & 1028.2 & 27.8 & 0.8 & \rm{yes}& 7.4\\
M87PN J123039.36+124259.0 & F1-F01  & 187.6640 & 12.7164 & 1539.2 & 27.8 & 1.0 & \rm{yes}& 6.6\\
M87PN J123025.34+124339.0 &   F1-F01  & 187.6056 & 12.7275 & 1397.8 & 27.6 & 1.0 & --& 7.5\\
M87PN J122948.74+122757.9 &F1-F02  & 187.4531 & 12.4661 & 1813.7 & 28.2 & 1.0 & \rm{yes}& 8.3\\
M87PN J123149.58+121132.2 &F2-F02  & 187.9566 & 12.1923 & 1230.7 & 28.4 & 1.0 & \rm{yes}& 6.7\\
M87PN J123133.76+120421.7 &F2-F02  & 187.8907 & 12.0727 & 933.5 & 27.7 & 0.5 & \rm{yes}& 6.6\\
M87PN J123139.12+120534.4 & F2-F02  & 187.9130 & 12.0929 & 1429.6 & 27.1 & 1.0 & \rm{yes}& 7.9\\
M87PN J122950.54+123717.0 &FEDGE  & 187.4606 & 12.6214 & 1116.2 & 28.0 & 1.0 & \rm{yes}& 5.6\\
M87PN J123128.41+120312.9 &F2-F01  & 187.8684 & 12.0536 & 1204.9 & 27.1 & 1.0 & \rm{yes}& 8.2\\
M87PN J123147.68+120818.2 &F2-F02  & 187.9487 & 12.1384 & 1687.2 & 26.4 & 1.0 & \rm{yes}& 7.4\\
M87PN J123138.04+120430.3 & F2-F02  & 187.9085 & 12.0751 & 1239.1 & 27.3 & 1.0 & \rm{yes}& 6.1\\
M87PN J123139.79+120456.2 &F2-F01  & 187.9158 & 12.0823 & 1323.5 & 27.3 & 1.0 & \rm{yes}& 5.5\\
M87PN J122946.89+123613.6 &F1-F03  & 187.4454 & 12.6038 & 1172.0 & 27.6 & 1.0 & \rm{yes}& 5.4\\
M87PN J122954.43+124052.6 &F1-F03  & 187.4768 & 12.6813 & 916.7 & 27.4 & 0.5 & \rm{yes}& 9.1\\
M87PN J123056.28+124229.1 & F1-F01  & 187.7345 & 12.7081 & 433.8 & 26.8 & 1.0 & \rm{yes}& 5.5\\
M87PN J123120.18+120147.2 &F2-F03  & 187.8341 & 12.0298 & 1495.5 & 27.7 & 1.0 & \rm{yes}& 7.0\\
M87PN J123121.55+120141.1 &F2-F03  & 187.8398 & 12.0281 & 1279.8 & 27.1 & 1.0 & \rm{yes}& 7.8\\
M87PN J122944.71+122744.2 &F1-F03  & 187.4363 & 12.4623 & 1414.0 & 26.9 & 1.0 & \rm{yes}& 4.2\\
M87PN J123154.96+120955.4 & F2-F02  & 187.9790 & 12.1654 & 1100.7 & 27.4 & 1.0 & \rm{yes}& 5.9\\
M87PN J122941.68+122929.7 &F1-F03  & 187.4237 & 12.4916 & 1065.3 & 27.5 & 0.9 & \rm{yes}& 10.8\\
M87PN J122941.64+123510.3 & F1-F02  & 187.4235 & 12.5862 & 1312.8 & 28.1 & 1.0 & \rm{yes}& 6.3\\
M87PN J123103.55+124227.0 &   F1-F01  & 187.7648 & 12.7075 & 1358.3 & 26.8 & 1.0 & \rm{yes}& 7.4\\
M87PN J122946.60+124027.1 &F1-F02  & 187.4442 & 12.6742 & 1737.6 & 27.2 & 1.0 & \rm{yes}& 7.9\\
M87PN J122946.58+124027.4 &FEDGE  & 187.4441 & 12.6743 & 1759.7 & 27.2 & 1.0 & \rm{yes}& 6.7\\
M87PN J122948.84+124211.5 & F1-F03  & 187.4535 & 12.7032 & 1146.2 & 27.2 & 1.0 & \rm{yes}& 5.7\\
M87PN J123143.58+120151.6 & F2-F01  & 187.9316 & 12.0310 & 1551.8 & 26.6 & 1.0 & \rm{yes}& 6.1\\
M87PN J122939.43+123826.1 &F1-F03  & 187.4143 & 12.6406 & 971.2 & 28.1 & 0.7 & \rm{yes}& 8.2\\
M87PN J122955.41+124530.2 &F1-F02  & 187.4809 & 12.7584 & 1314.6 & 27.4 & 1.0 & \rm{yes}& 7.9\\
M87PN J123101.10+124450.6 &F1-F01  & 187.7546 & 12.7474 & 1816.1 & 27.5 & 1.0 & \rm{yes}& 8.7\\
M87PN J123019.80+120549.9 &  F2-F03  & 187.5825 & 12.0972 & 1521.9 & 27.0 & 1.0 & \rm{yes}& 8.6\\
M87PN J123142.21+115955.3 &F2-F01  & 187.9259 & 11.9987 & 1096.5 & 27.3 & 1.0 & \rm{yes}& 6.9\\
M87PN J122936.48+123917.6 & F1-F03  & 187.4020 & 12.6549 & 947.9 & 27.6 & 0.6 & \rm{yes}& 8.7\\
M87PN J122952.05+124614.1 &FEDGE  & 187.4669 & 12.7706 & 942.5 & 27.5 & 0.6 & \rm{yes}& 6.3\\
M87PN J123033.45+120232.2 &F2-F03  & 187.6394 & 12.0423 & 1180.4 & 27.5 & 1.0 & \rm{yes}& 6.3\\
M87PN J123050.20+115944.8 &F2-F03  & 187.7092 & 11.9958 & 1409.2 & 27.8 & 1.0 & \rm{yes}& 7.7\\
M87PN J123047.20+124722.5 &F1-F01  & 187.6967 & 12.7896 & 1260.0 & 28.8 & 1.0 & \rm{yes}& 6.6\\
M87PN J123044.66+120022.3 &F2-F03  & 187.6861 & 12.0062 & 1252.9 & 28.3 & 1.0 & \rm{yes}& 8.2\\
M87PN J123106.04+115800.4 &F2-F03  & 187.7752 & 11.9668 & 1254.0 & 27.1 & 1.0 & \rm{yes}& 9.0\\
M87PN J122928.80+123309.7 &  F1-F02  & 187.3700 & 12.5527 & 1152.2 & 28.2 & 1.0 & \rm{yes}& 8.7\\
M87PN J123210.24+120823.6 &F2-F02  & 188.0427 & 12.1399 & 1412.2 & 27.5 & 1.0 & \rm{yes}& 5.6\\
M87PN J122927.79+122928.6 &F1-F03  & 187.3658 & 12.4913 & 757.3 & 26.9 & 1.0 & \rm{yes}& 7.1\\
M87PN J122926.32+123645.7 &FEDGE  & 187.3597 & 12.6127 & 1260.0 & 27.9 & 1.0 & \rm{yes}& 6.7\\
M87PN J122930.93+122348.8 &F1-F03  & 187.3789 & 12.3969 & 1290.6 & 29.2 & 1.0 & \rm{yes}& 7.5\\
M87PN J123117.04+124447.0 & F1-F01  & 187.8210 & 12.7464 & 1214.5 & 28.3 & 1.0 & \rm{yes}& 5.6\\
M87PN J123212.76+120637.4 &F2-F02  & 188.0532 & 12.1104 & 1515.3 & 29.0 & 1.0 & \rm{yes}& 5.7\\
M87PN J122925.92+122758.6 & F1-F03  & 187.3580 & 12.4663 & 1739.3 & 26.7 & 1.0 & --$^{b}$& 4.4 \\
M87PN J123123.56+115421.9 &F2-F03  & 187.8482 & 11.9061 & 1405.6 & 27.1 & 1.0 & \rm{yes}& 5.6\\
M87PN J123140.27+115202.6 &F2-F01  & 187.9178 & 11.8674 & 1422.4 & 27.6 & 1.0 & \rm{yes}& 8.1\\
M87PN J123100.84+121703.8 & FC* & 187.7535 & 12.2844 & 807.1 & 27.6 & 0.1 & \rm{yes}& 6.5\\
M87PN J123053.85+121621.0 &   FC* & 187.7244 & 12.2725 & 793.9 & 26.8 & 0.1 & \rm{yes}& 7.4\\
M87PN J123100.50+121516.2 & FC* & 187.7521 & 12.2545 & 759.7 & 26.8 & 0.2 & \rm{yes}& 8.2\\
M87PN J123032.01+123118.8 &F1-F03*  & 187.6334 & 12.5219 & 906.5 & 27.7 & 0.1 & \rm{yes}& 8.0\\
M87PN J123027.50+123029.5 &F1-F03* & 187.6146 & 12.5082 & 822.6 & 27.6 & 0.1 & \rm{yes}& 6.8\\
M87PN J123051.84+123125.3 & F1-F01*  & 187.7160 & 12.5237 & 813.7 & 27.8 & 0.1 & \rm{yes}& 6.5\\
M87PN J123039.50+123256.4 & F1-F01*  & 187.6646 & 12.5490 & 912.5 & 27.9 & 0.1 & \rm{yes}& 8.8\\
M87PN J123037.17+123306.1 &F1-F03* & 187.6549 & 12.5517 & 871.2 & 27.6 & 0.1 & \rm{yes}& 7.9\\
M87PN J123052.03+123249.2 & F1-F01*  & 187.7168 & 12.5470 & 928.7 & 28.1 & 0.2 & \rm{yes}& 8.7\\
M87PN J123109.62+123210.6 & FC* & 187.7901 & 12.5363 & 1060.5 & 28.1 & 0.2 & \rm{yes}& 5.3\\
\hline \\

VICPN J123034.68+121011.2 & FC & 187.6445 & 12.1698 & 1888.6 & 28.2 & 0.1 & --& 5.8\\
VICPN J123032.16+120937.0 & F2-F03  & 187.6340 & 12.1603 & 1911.9 & 27.4 & 0.1 & \rm{yes}& 7.9\\
VICPN J123026.49+120209.6 & F2-F03  & 187.6104 & 12.0360 & 1762.7 & 27.4 & 0.1 & \rm{yes}& 7.5\\
VICPN J123025.29+120140.0 &F2-F03  & 187.6054 & 12.0278 & 1718.4 & 27.2 & 0.1 & \rm{yes}& 7.2\\
VICPN J123042.36+122538.2 & FC & 187.6765 & 12.4273 & 795.7 & 27.1 & 0.2 & \rm{yes}& 6.2\\
VICPN J123100.86+122302.7 &FC & 187.7536 & 12.3841 & 1080.3 & 27.5 & 0.5 & \rm{yes}& 6.1\\
VICPN J123054.62+122608.5 &FC & 187.7276 & 12.4357 & 1032.3 & 27.1 & 0.4 & \rm{yes}& 8.5\\
VICPN J123034.63+122554.1 &FC & 187.6443 & 12.4317 & 944.9 & 27.0 & 0.4 & --& 6.1\\
VICPN J123103.72+121949.8 &  FC & 187.7655 & 12.3305 & 891.5 & 26.7 & 0.4 & \rm{yes}& 7.1\\
VICPN J123048.24+121833.4 & FC & 187.7010 & 12.3093 & 913.7 & 26.6 & 0.4 & \rm{yes}& 6.5\\
VICPN J123053.59+122750.7 &FC & 187.7233 & 12.4641 & 660.9 & 27.2 & 0.3 & \rm{yes}& 6.3\\
VICPN J123031.41+122505.1 &F1-F03  & 187.6309 & 12.4181 & 684.2 & 27.1 & 0.2 & \rm{yes}& 8.9\\
VICPN J123104.56+122458.3 & FC & 187.7690 & 12.4162 & 1042.5 & 27.2 & 0.4 & --& 5.8\\
VICPN J123102.49+122711.8 &FC & 187.7604 & 12.4533 & 1088.1 & 26.9 & 0.5 & \rm{yes}& 7.6\\
VICPN J123041.30+123226.1 &F1-F01  & 187.6721 & 12.5406 & 1309.2 & 27.1 & 0.4 & \rm{yes}& 8.9\\
VICPN J123025.41+123034.5 &F1-F03  & 187.6059 & 12.5096 & 976.0 & 26.8 & 0.3 & \rm{yes}& 8.5\\
VICPN J123111.28+121310.2 &  F2-F02  & 187.7970 & 12.2195 & 829.8 & 27.0 & 0.3 &-- & 9.0\\
VICPN J123043.36+123504.2 & F1-F01  & 187.6807 & 12.5845 & 872.4 & 28.0 & 0.5 & \rm{yes}& 8.4\\
VICPN J123014.97+122214.1 &F1-F03  & 187.5624 & 12.3706 & 1043.2 & 27.3 & 0.2 & \rm{yes}& 5.1\\
VICPN J123116.65+121017.4 & F2-F02  & 187.8194 & 12.1715 & 882.0 & 27.4 & 0.4 & --& 8.2\\
VICPN J123116.92+120329.1 & F2-F03  & 187.8205 & 12.0581 & 746.0 & 27.0 & 0.5 & \rm{yes}& 7.9\\
VICPN J123033.48+124330.3 & FEDGE  & 187.6395 & 12.7251 & 741.8 & 26.6 & 0.5 & \rm{yes}& 5.8\\
VICPN J122948.45+123300.0 &  F1-F03  & 187.4519 & 12.5500 & 844.8 & 27.3 & 0.4 & \rm{yes}& 3.0\\
VICPN J122959.28+124216.9 & F1-F03  & 187.4970 & 12.7047 & 763.9 & 26.9 & 0.4 & \rm{yes}& 7.6\\
VICPN J122955.22+124242.4 &FEDGE  & 187.4801 & 12.7118 & 717.8 & 28.5 & 0.5 & \rm{yes}& 5.8\\
VICPN J122940.44+123515.3 & F1-F03  & 187.4185 & 12.5876 & 744.2 & 26.5 & 0.5 & \rm{yes}& 4.1\\
VICPN J122947.66+124251.4 &F1-F02  & 187.4486 & 12.7143 & 881.4 & 27.0 & 0.5 & \rm{yes}& 10.0\\
VICPN J123122.77+124545.7 &F1-F01  & 187.8449 & 12.7627 & 1049.7 & 27.8 & 0.4 & \rm{yes}& 6.5\\

\hline \\
VICPN J123025.896+123050.0 & F1-F03  & 187.6079 & 12.5140 & -194.7 & 28.9 & 0.0 & \rm{yes}& 7.3\\
VICPN J123053.688+121322.4 &FC & 187.7237 & 12.2229 & 2390.6 & 27.7 & 0.0 & \rm{yes}& 5.6\\
VICPN J123039.648+121449.5 &FC & 187.6652 & 12.2471 & -1048.5 & 27.7 & 0.0 & --& 6.0\\
VICPN J123119.512+121416.8 & F2-F02  & 187.8313 & 12.2380 & 299.6 & 28.1 & 0.0 & \rm{yes}& 6.5\\
VICPN J123123.136+122358.2 & FC & 187.8464 & 12.3995 & 31.2 & 26.6 & 0.0 & \rm{yes}& 5.7\\
VICPN J123115.288+121034.6 &F2-F03  & 187.8137 & 12.1763 & -2005.4 & 26.3 & 0.0 & --& 4.1\\
VICPN J123043.632+123828.3 &F1-F01  & 187.6818 & 12.6412 & -187.5 & 27.0 & 0.0 & \rm{yes}& 5.2\\
VICPN J123122.920+122942.7 & F1-F01  & 187.8455 & 12.4952 & 254.6 & 26.9 & 0.0 & \rm{yes}& 4.3\\
VICPN J123021.000+123918.7 &    F1-F02  & 187.5875 & 12.6552 & -343.9 & 26.8 & 0.0 & \rm{yes}& 8.5\\
VICPN J123005.472+123545.9 &FEDGE  & 187.5228 & 12.5961 & -441.0 & 27.2 & 0.0 & --& 6.1\\
VICPN J123038.448+120746.5 &F2-F03  & 187.6602 & 12.1296 & -452.4 & 27.9 & 0.0 & \rm{yes}& 6.4\\
VICPN J123114.808+123637.8 & F1-F01  & 187.8117 & 12.6105 & 2306.8 & 27.1 & 0.0 & \rm{yes}& 9.6\\
VICPN J123126.496+123253.8 &F1-F01  & 187.8604 & 12.5483 & -272.0 & 28.7 & 0.0 & --& 4.5\\
VICPN J123145.912+120949.3 &F2-F02  & 187.9413 & 12.1637 & 166.6 & 27.4 & 0.0 & \rm{yes}& 7.8\\
VICPN J123038.424+124412.1 &FEDGE  & 187.6601 & 12.7367 & -399.6 & 27.0 & 0.0 & \rm{yes}& 5.5\\
VICPN J123140.536+120443.3 &F2-F01  & 187.9189 & 12.0787 & -2949.1 & 29.6 & 0.0 & --$^{a}$& 6.3\\
VICPN J123143.632+120526.8 &F2-F01  & 187.9318 & 12.0908 & -379.3 & 28.3 & 0.0 &-- & 5.7\\
VICPN J123127.648+120202.4 & F2-F03  & 187.8652 & 12.0340 & -186.3 & 28.4 & 0.0 & --& 6.2\\
VICPN J122948.576+124053.4 & F1-F02  & 187.4524 & 12.6815 &  2566.2 & 28.5 & 0.0 & \rm{yes}& 6.7\\
VICPN J123118.672+123918.3 &F1-F01  & 187.8278 & 12.6551 & -1235.5 & 27.7 & 0.0 & \rm{yes}& 5.1\\
VICPN J122937.296+123536.2 &FEDGE  & 187.4054 & 12.5934 & -572.8 & 27.1 & 0.0 & \rm{yes}& 6.5\\
VICPN J122937.056+123001.0 &F1-F03  & 187.4044 & 12.5003 & -193.5 & 27.8 & 0.0 & \rm{yes}& 8.3\\
VICPN J123143.632+120125.3 &F2-F02  & 187.9318 & 12.0237 & -91.1 & 26.9 & 0.0 & \rm{yes}& 6.8\\
VICPN J122942.480+122254.1 & F1-F03  & 187.4270 & 12.3817 & 234.3 & 32.0 & 0.0 & \rm{yes}& 7.2\\
VICPN J123153.520+120333.8 & F2-F01  & 187.9730 & 12.0594 & 2277.4 & 26.9 & 0.0 & \rm{yes}& 4.3\\
VICPN J123034.344+120252.4 &F2-F03  & 187.6431 & 12.0479 & 2412.8 & 26.6 & 0.0 & \rm{yes}& 4.0\\
VICPN J123147.976+120006.4 &F2-F02  & 187.9499 & 12.0018 & 2371.5 & 28.9 & 0.0 & \rm{yes}& 6.5\\
VICPN J123110.296+124541.7 &F1-F01  & 187.7929 & 12.7616 & 2479.3 & 27.7 & 0.0 & \rm{yes}& 7.6\\
VICPN J123101.200+115619.3 &  F2-F03  & 187.7550 & 11.9387 & 205.5 & 27.2 & 0.0 & \rm{yes}& 4.6\\
VICPN J123214.064+121501.4 &F2-F02  & 188.0586 & 12.2504 & -3227.7 & 27.6 & 0.0 & --$^{a}$& 6.0\\
VICPN J123213.224+120737.2 & F2-F02  & 188.0551 & 12.1270 & -446.4 & 28.8 & 0.0 & \rm{yes}& 6.0\\
VICPN J122923.232+123354.0 &   FEDGE  & 187.3468 & 12.5650 & -988.0 & 28.8 & 0.0 & \rm{yes}& 5.3\\
VICPN J122922.368+123655.8 & FEDGE  & 187.3432 & 12.6155 & 2231.9 & 29.9 & 0.0 & \rm{yes}& 6.1\\
VICPN J123147.328+115509.8 &F2-F01  & 187.9472 & 11.9194 & 2279.2 & 26.2 & 0.0 & \rm{yes}& 4.5\\
VICPN J123230.576+121523.0 &F2-F02  & 188.1274 & 12.2564 & -3462.6 & 28.6 & 0.0 & --$^{a}$&  4.5\\
VICPN J123156.712+121111.7 &F2-F02  & 187.9863 & 12.1866 & 408.0 & 26.4 & 0.0 & \rm{yes}& 4.7\\
VICPN J123011.328+123611.8 &F1-F03  & 187.5472 & 12.6033 & 2117.4 & 27.7 & 0.0 & \rm{yes}& 7.7\\
VICPN J123158.944+121029.6 &F2-F02  & 187.9956 & 12.1749 & 387.7 & 27.9 & 0.0 & \rm{yes}& 7.7\\
VICPN J123036.648+122943.4 &F1-F02  & 187.6527 & 12.4954 & 584.2 & 27.0 & 0.0 & \rm{yes}& 7.9\\
VICPN J123100.384+124254.3 &F1-F01  & 187.7516 & 12.7151 & 2179.7 & 28.1 & 0.0 & \rm{yes}& 7.1\\
VICPN J123158.080+120938.8 & F2-F02  & 187.9920 & 12.1608 & 352.3 & 27.1 & 0.0 & \rm{yes}& 7.3\\
VICPN J123138.016+121015.2 &F2-F02  & 187.9084 & 12.1709 & 2209.1 & 26.6 & 0.0 & --& 4.5\\
VICPN J123013.272+124231.6 &F1-F02  & 187.5553 & 12.7088 & 2231.3 & 27.9 & 0.0 & \rm{yes}& 6.3\\
VICPN J123156.304+120958.3 &F2-F02  & 187.9846 & 12.1662 & 315.1 & 26.8 & 0.0 & \rm{yes}& 7.6\\
VICPN J122905.832+123904.6 &F71    & 187.2743 & 12.6513 & 2197.3 &26.8 & 0.0& \rm{yes}& 15.0\\

\end{longtable}
\tablefoot{
\tablefoottext{a}{The Doppler-shifted [OIII]]$\lambda$4959\AA\ emission line falls at a shorter wavelength than the blue edge of the FLAMES filter.}\\
\tablefoottext{b}{Spectrum deprecated in the wavelength region around the Doppler-shifted [OIII]]$\lambda$4959\AA\ emission line.}}
\\
\end{appendix}

\begin{acknowledgements}
  AL thanks Jonathan N. Elliott for advices and suggestions on the manuscript and Guido Avvisati who
  helped in the installation of IDL libraries. OG thanks Natalya Lyskova and Eugene Churazov for discussions on the M87
  X-ray potential. AL and MAR wish to acknowledge ESO support via the Science Support Discretionary Funds 18/28. This research has made use of the NASA/IPAC Extragalactic Database (NED) which is
  operated by the Jet Propulsion Laboratory, California Institute of Technology, under contract with
  the National Aeronautics and Space Administration.
\end{acknowledgements}

\bibliographystyle{aa}
\bibliography{PNrefs_2,PNrefs,PNrefs_chevron}
\newpage
\newpage
\end{document}